\documentclass[twocolumn,showpacs,preprintnumbers,amsmath,amssymb,floatfix]{revtex4-1}

\usepackage{graphicx}
\usepackage{dcolumn}
\usepackage{bm}
\usepackage{epsfig}
\usepackage{epstopdf}
\usepackage{subfigure}

\begin{document}
\title{Scaling behavior of explosive percolation on the square lattice\\ }
\author{Robert M. Ziff}%
\email{rziff@umich.edu}
\affiliation{%
Michigan Center for Theoretical Physics and Department of Chemical Engineering, University of Michigan, Ann
Arbor MI 48109-2136.
}%

\date{\today}%

\begin{abstract}
Clusters generated by the product-rule growth model of Achlioptas, D'Souza, and Spencer on a two-dimensional
square lattice are shown to obey qualitatively different scaling behavior than standard (random growth) percolation.  The threshold with unrestricted bond placement (allowing loops)
is found precisely using several different criteria based upon both moments and wrapping probabilities,
yielding $p_c = 0.526565 \pm 0.000005$, consistent
with the recent result of Radicchi and Fortunato.  The correlation-length exponent $\nu$ is found to be close to 1.  The qualitative difference from regular percolation is shown dramatically in the behavior of the percolation
probability $P_\infty$ (size of largest cluster), the susceptibility, and of the second moment of finite clusters,
where discontinuities appears at the threshold.
The critical cluster-size distribution does not follow a consistent power-law for the range of system sizes we study ($L \le 8192$)
but may approach a power-law with $\tau > 2$ for larger $L$.
\end{abstract}

\maketitle

\section{\label{sec:Introduction}Introduction\protect\\ }

Recently, there has been a great deal of interest in a  model of ``explosive growth" of percolation clusters by the
so-called Achlioptas process \cite{AchlioptasDSouzaSpencer09}, in which 
two randomly chosen unoccupied bonds in a system are examined, and the bond that minimizes
the product of the size of the two clusters to which it is attached becomes the next one to occupied.
This procedure, called the product rule (PR) \cite{AchlioptasDSouzaSpencer09},
was originally studied on Erd\H os-R\'enyi random graphs \cite{AchlioptasDSouzaSpencer09}, then on two-dimensional square lattices \cite{Ziff09}
and scale-free networks \cite{RadicchiFortunato09,ChoKimParkKahngKim09}.  
Other recent papers on explosive and biased percolation include
\cite{FriedmanLandsberg09,RozenfeldGallosMakse09,ChoKahngKim10,MoreiraOliveiraReisHerrmannAndrade09,MannaChatterjee09,HooyberghsEtAl10,DSouzaMitzenmacher10,ChoKimNohKahngKim10,AldousOngZhou10,RadicchiFortunato10,AraujoHerrmann10,
BasuBasuKunduMohanty10}.
Interest in this process derives from
its unusual explosive behavior, suggesting a first-order transition, with apparent discontinuities in several quantities.  Many of its properties have yet to be discovered.

\begin{figure*}[htbp] 
   \centering
   \includegraphics[width=3in]{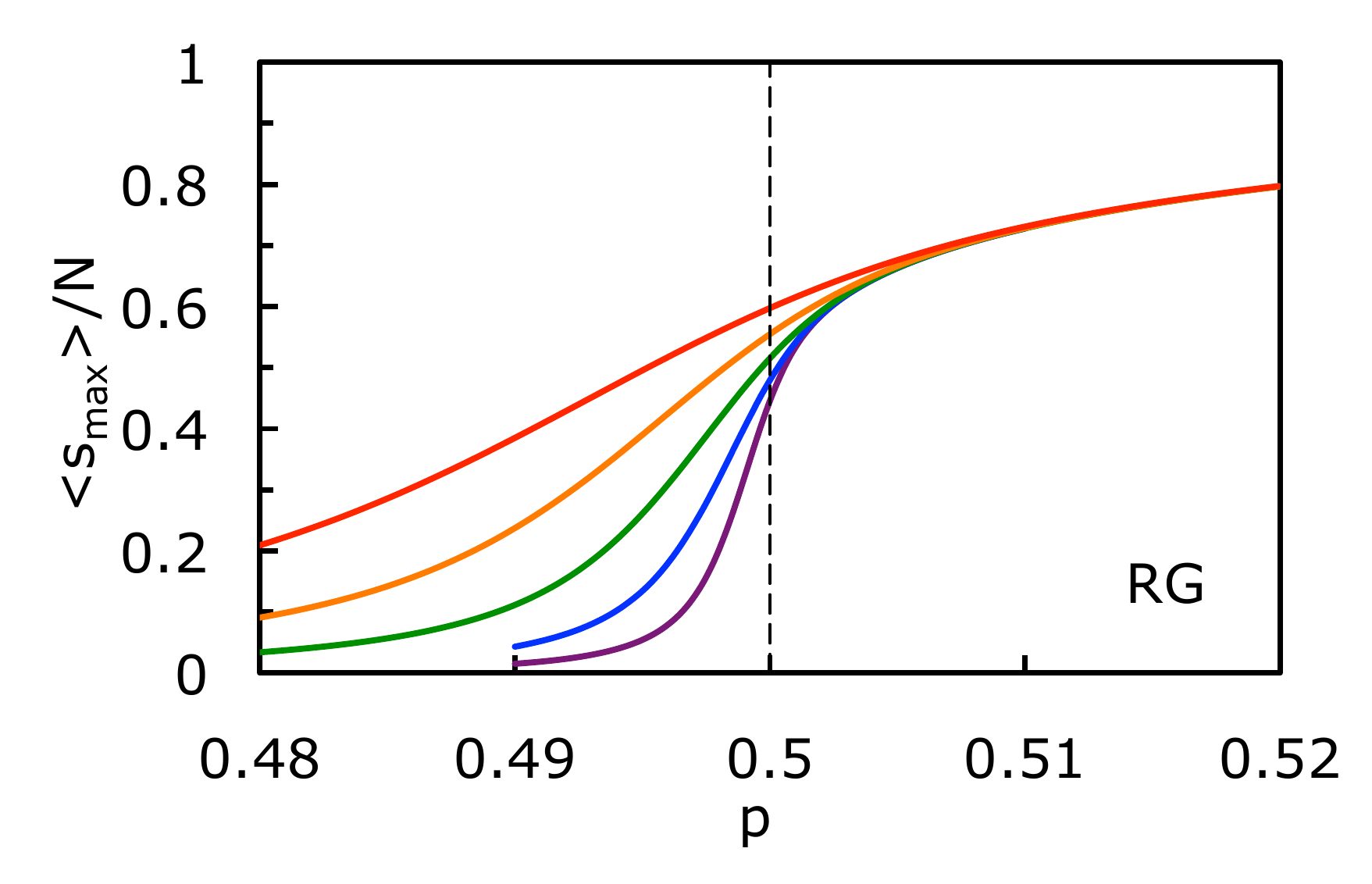} 
   \includegraphics[width=3in]{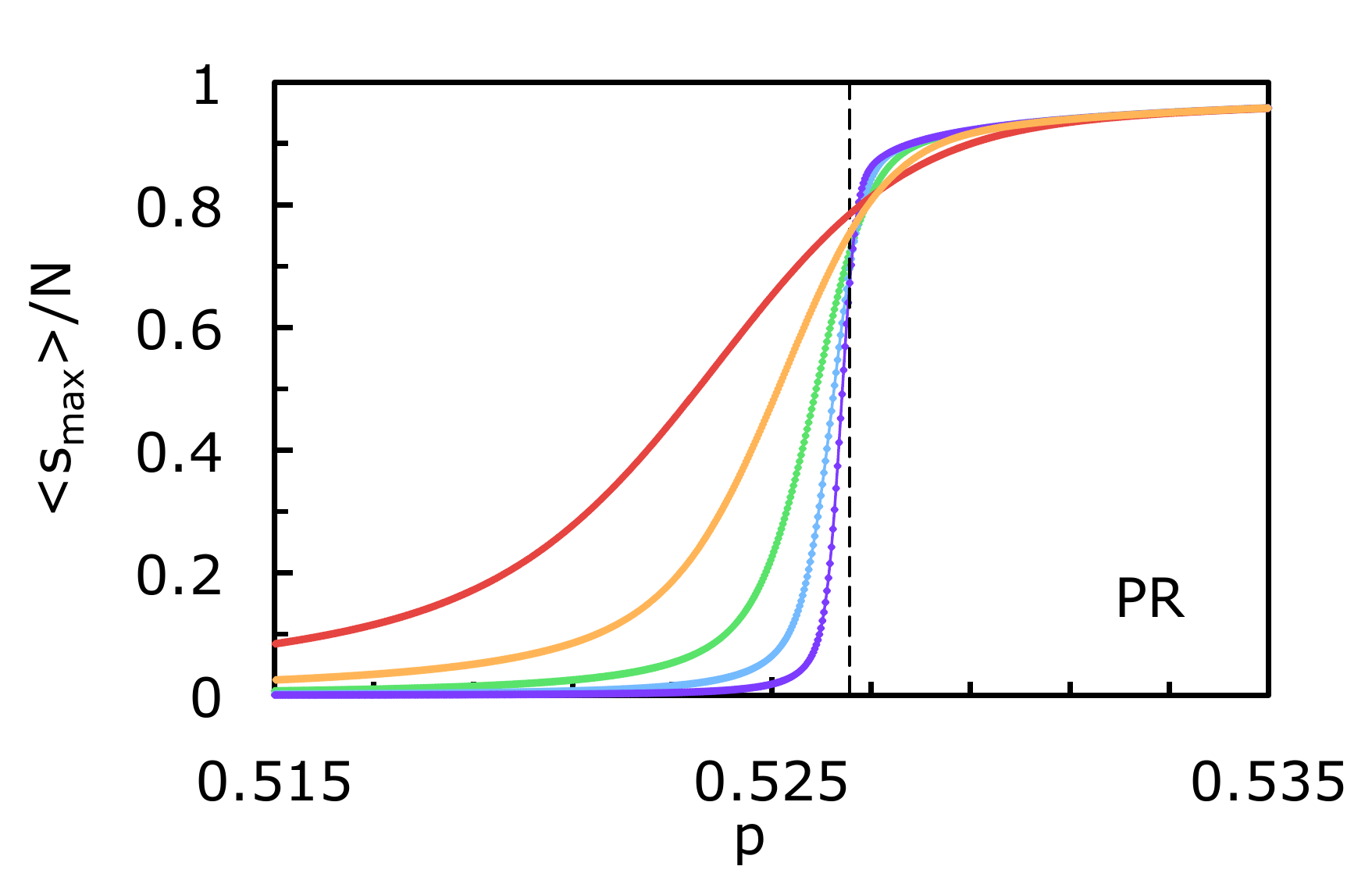} 
   \caption{(Color online) $\langle s_\mathrm{max} \rangle/N$ (or $P_\infty$) vs.\ $p$ as a function of system size.   Here, as in all of the figures, we show  curves for $L = 128$ (red), 256 (orange), 512 (green), 1024 (blue), and 2048 (violet) --- in general,
   from more gradual ($L = 128$) to sharper ($L=2048$).  Vertical dashed lines show the transition points, $p_c = 0.5$ (RG) and 
   $p_c = 0.526565$ (PR).  Plots for RG are always shown on the left, and those for PR are shown on the right.
   The scaling behavior of $\langle s_\mathrm{max} \rangle/N$ at $p_c$ is shown in Fig.\ \ref{fig:M2smax}.}
   \label{fig:smax}
\end{figure*}

In this paper we examine in more detail the PR model on the regular square lattice,
especially in regards to how it differs from random growth (RG), in which bonds
are added one at a time, and which corresponds
to standard percolation.

Some preliminary results were given in \cite{Ziff09}, where the 
width of the distribution $\Delta/N$ was investigated.  As in \cite{AchlioptasDSouzaSpencer09},
$\Delta/N$ was defined as the
difference in times in which the maximum cluster size $s_\mathrm{max}$ goes from
$\sqrt{N}$ to $0.5 N$, where $N$ is the number of sites.  (Here time is identical to the number of bonds added.) 
For the Erd\H os-R\'enyi graph, the authors of \cite{AchlioptasDSouzaSpencer09} found that $\Delta/N \to 0$ as $N \to \infty$ for the PR model, while $\Delta/N \to$ const.\ for 
the RG model, showing that the two transitions are qualitatively different.
In \cite{Ziff09} it was found that for the square lattice, on the other hand, $\Delta/N \to 0$
for both the PR and RG models, but with different powers in $N$.  
It however turns out that if a larger (and in the 
case of the square boundary, more appropriate) criterion for the upper end of the gap $\Delta$ were used,
say $s_\mathrm{max} = 0.7 N$, then 
indeed one would find that $\Delta/N \to$ const.\ as $N \to \infty$ for the RG model and still to zero for the PR model.
So with this criterion, the two models are qualitatively different on the square lattice just as for the Erd\H os-R\'enyi graph.
(Even with a criterion of $s_\mathrm{max} = 0.5 N$, one should have $\Delta/N \to$ const.\ for the RG model on the square lattice, but one would
have to go to a very large system to see it.)

\begin{figure*}[htbp] 
   \centering
   \includegraphics[width=3in]{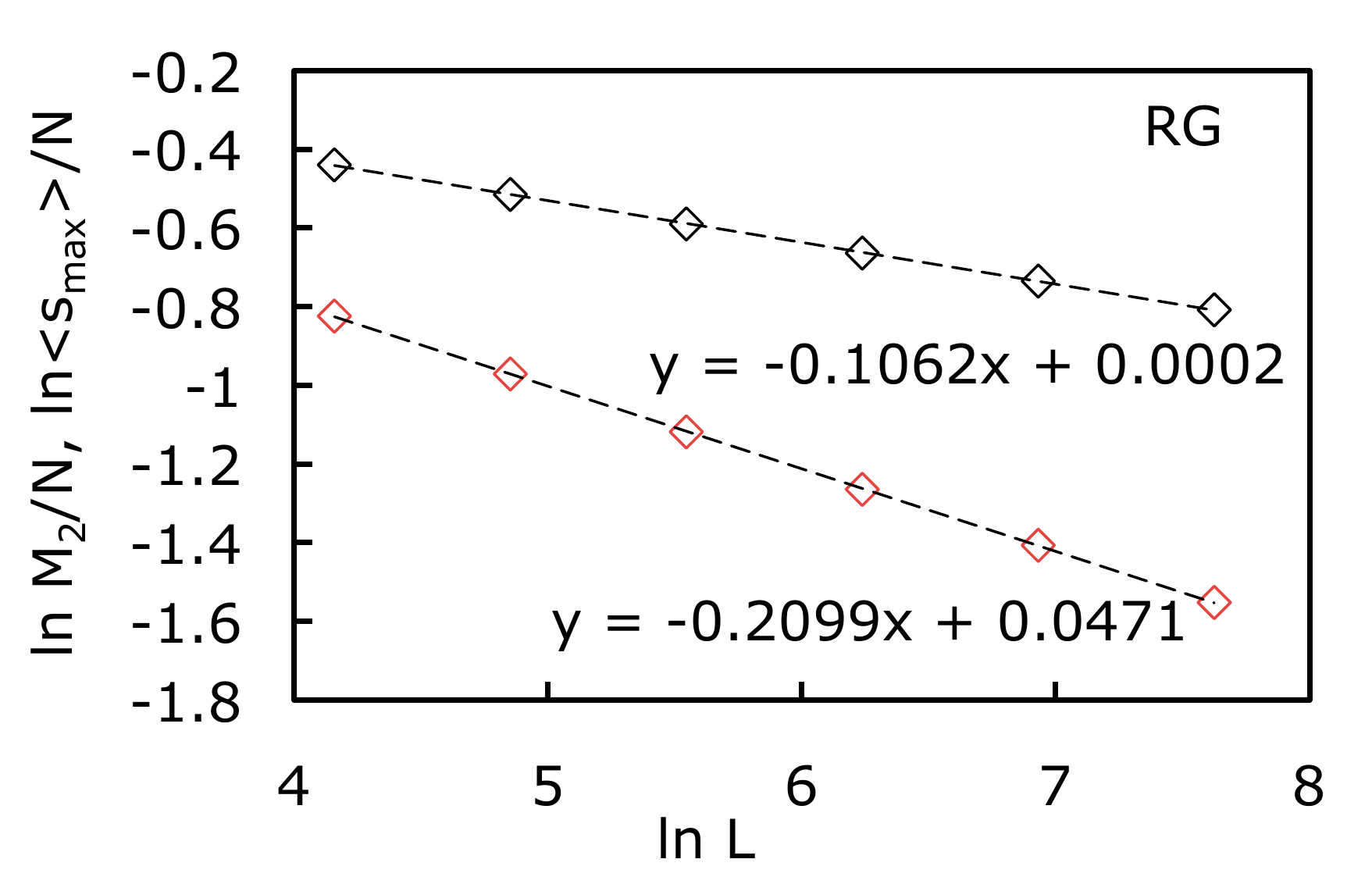} 
   \includegraphics[width=3in]{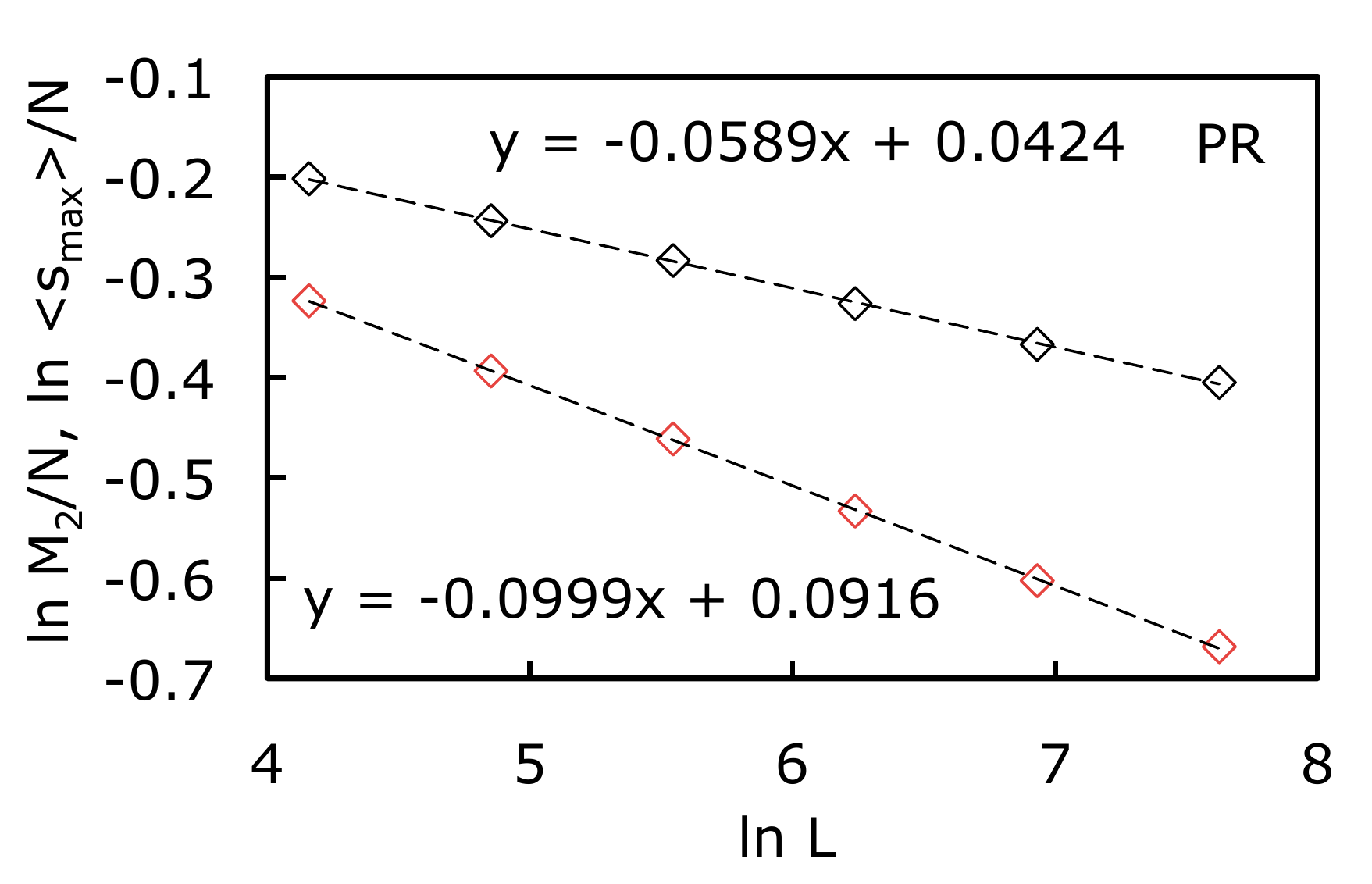}
   \caption{(Color online) $\ln (\langle s_\mathrm{max}(p_c) \rangle /N)$  (upper points) and $ \ln (M_2(p_c)/N)$  (lower points) as a function of $\ln L$, where $p_c = 0.5$ (RG) and $p_c = 0.526565$ (PR).  
  Linear fits to the points are shown on the plots, where $y$ represents the the abscissa value and $x$ represents $\ln L$. In these plots, we have also included  data for runs at $L = 64$.}
   \label{fig:M2smax}
\end{figure*}

Recently, Radicchi and Fortunato have also studied both the PR and RG models on the square lattice,
and analyzed their behavior in the context of standard two-parameter scaling \cite{RadicchiFortunato10}.
However, as they mention, it is unclear in what sense
this scaling can be applied to the PR problem, considering
that several of the quantities show discontinuities.  In this paper, we consider the behavior of 
wrapping probabilities
as well as quantities related to the size distribution such as moments.
The former refers to having a cluster that connects around the
toroidal boundaries of the periodic system (the torus), and is the analog of crossing probabilities
for open systems.  In order to study scaling behavior precisely,
it is also necessary to know the transition point precisely, and we determine it here
using a variety of methods.  While the convergence of various estimates in ordinary percolation is
well-known \cite{StaufferAharony94,ZiffNewman02}, that is not the case for the PR model, so
the convergence behavior is also studied.

\section{Procedure}
Actually, there is a subtle but significant difference in the treatment of the PR process considered previously
by the present author \cite{Ziff09} and that by Radicchi and Fortunato
\cite{RadicchiFortunato10}. 
In \cite{Ziff09}, it was assumed that bonds could only be added between different clusters.
This assumption bypassed the question of how to assign weights when a bond connects sites that are
part of the same cluster, and for the RG model corresponds to ``loop-less" percolation previously considered
by Manna and Subramanian \cite{MannaSubramanian96}.   On the other hand, in   \cite{RadicchiFortunato10} the authors
considered that bonds can be placed anywhere, including within the same cluster. 
Thus, there is a difference in the scaling of the time, but also a difference
in the weights with which new bonds are added, so these two processes are not equivalent.

In this paper, we follow the unrestricted
bond placement convention used by Radicchi and Fortunato in \cite{RadicchiFortunato10}.  When an internal bond is selected, we use for its weight the square of the size of the
cluster it is part of.  We characterize the size of a cluster (or component) by the number of sites it contains.  

To carry out these simulations, we used the algorithm of Newman and Ziff \cite{NewmanZiff00,NewmanZiff01} in which clusters
are represented as a tree and a union-find algorithm (modified for
the PR) is used to join clusters together.  A randomly ordered list of all bonds is made initially,
and bonds are taken off that list in pairs.  The bond that is not selected according to the PR is put back on the
list randomly by  switching with a randomly chosen member remaining on the list.  We also considered 
the less efficient
procedure of not using a bond list but just randomly selecting bonds on the lattice,
skipping over those that were already chosen until two free ones were found.  Both methods led to identical  results.

To determine cluster wrapping, we assigned extra variables ``xcoor" and ``ycoor" to each lattice site.
These quantities are the $x-$ and $y-$ coordinates of that site with respect to the first site of the cluster, without adjusting
for the periodic boundary conditions.  Wrapping is then indicated when an intra-cluster bond
leads to a difference in a coordinate by a multiple of the lattice width or height 
\cite{MachtaChoiLuckeSchweizerChayes96}.

The algorithm of \cite{NewmanZiff00} allows one to find the various quantities for all values of $p$ in one simulation.
We did not carry out the convolution step with a binomial distribution to get the grand canonical
(fixed probability) rather than canonical (fixed number) results, as the differences between the two ensembles
for the systems we studied are small.
We everywhere consider square lattices and square boundaries, with $N = L\times L$ sites and periodic b.c.
Many runs were made to get good statistics, ranging from $1\,000\,000$ runs for $L = 128$ to $150\,000$ runs
for $L = 2048$.  In general, the number of runs was sufficient so that the errors are smaller
than the symbols or width of the lines or symbols we used to plot the results.
We also considered runs for $L = 8192$ for measuring the size distribution.

\begin{figure*}[htbp] 
   \centering
   \includegraphics[width=3in]{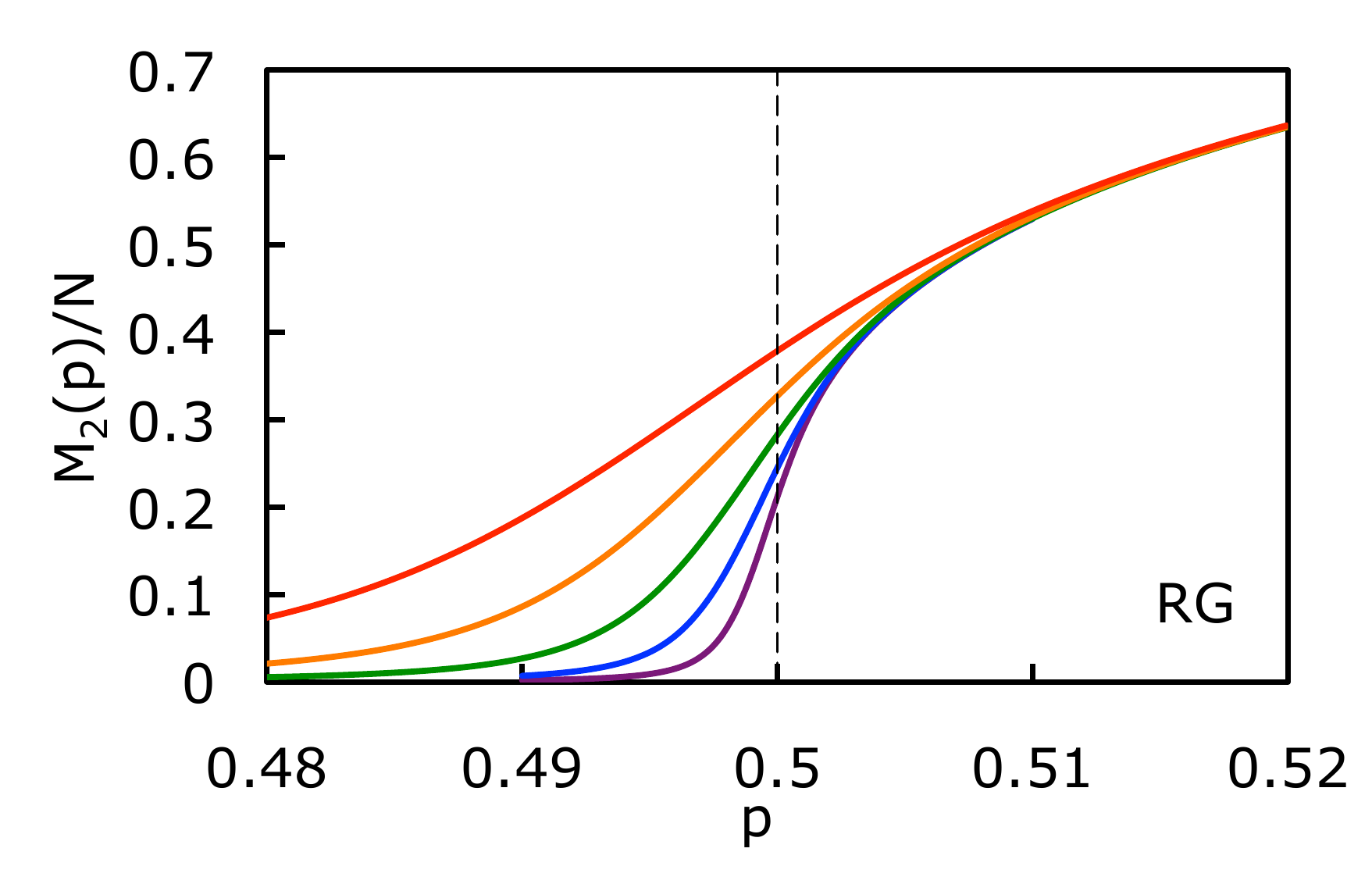} 
   \includegraphics[width=3in]{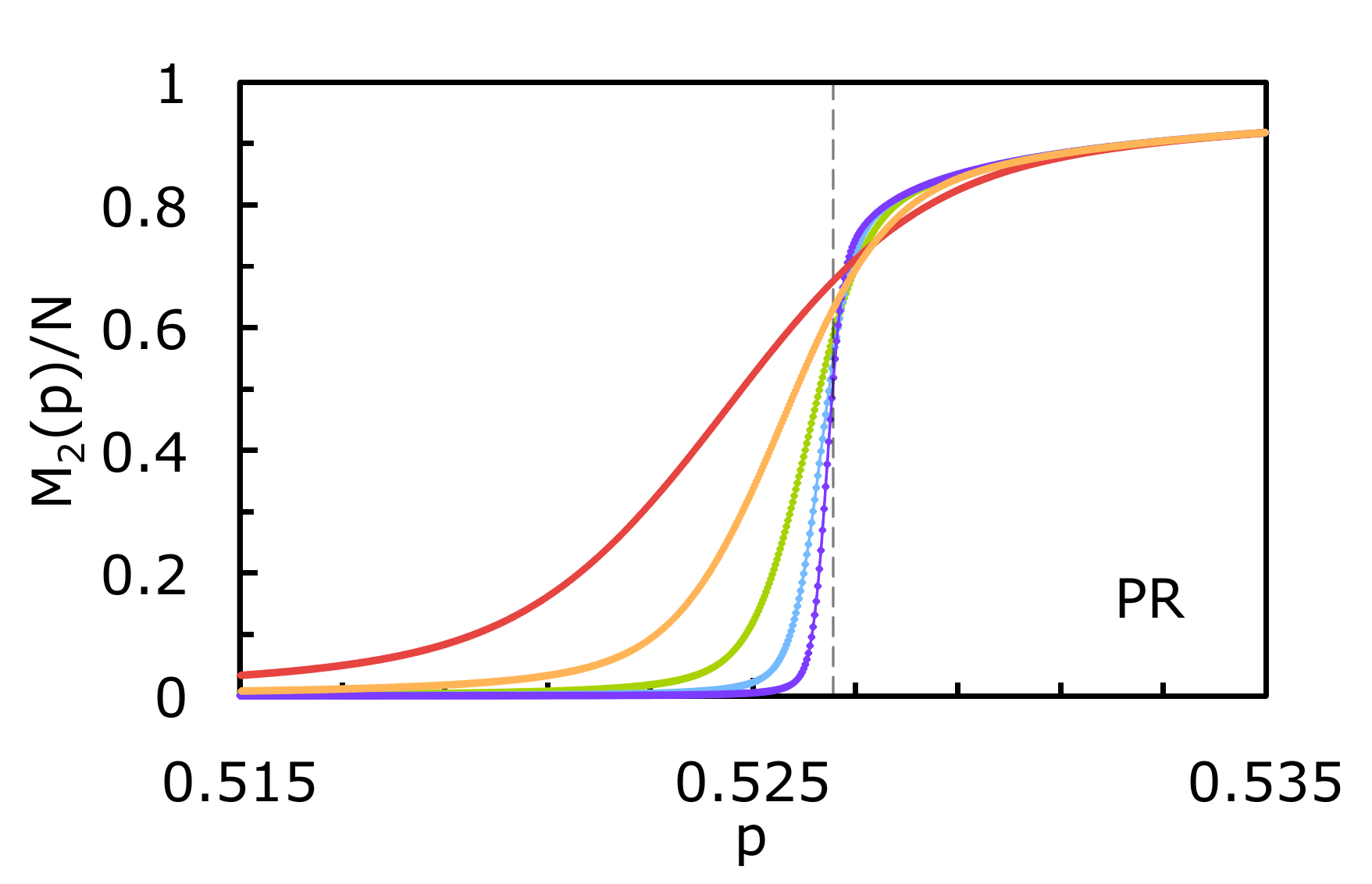} 
   \caption{(Color online) Scaled second moment $M_2(p)/N$ as a function of $p$.}
   \label{fig:M2}
\end{figure*}

\begin{figure*}[htbp] 
   \centering
   \includegraphics[width=3in]{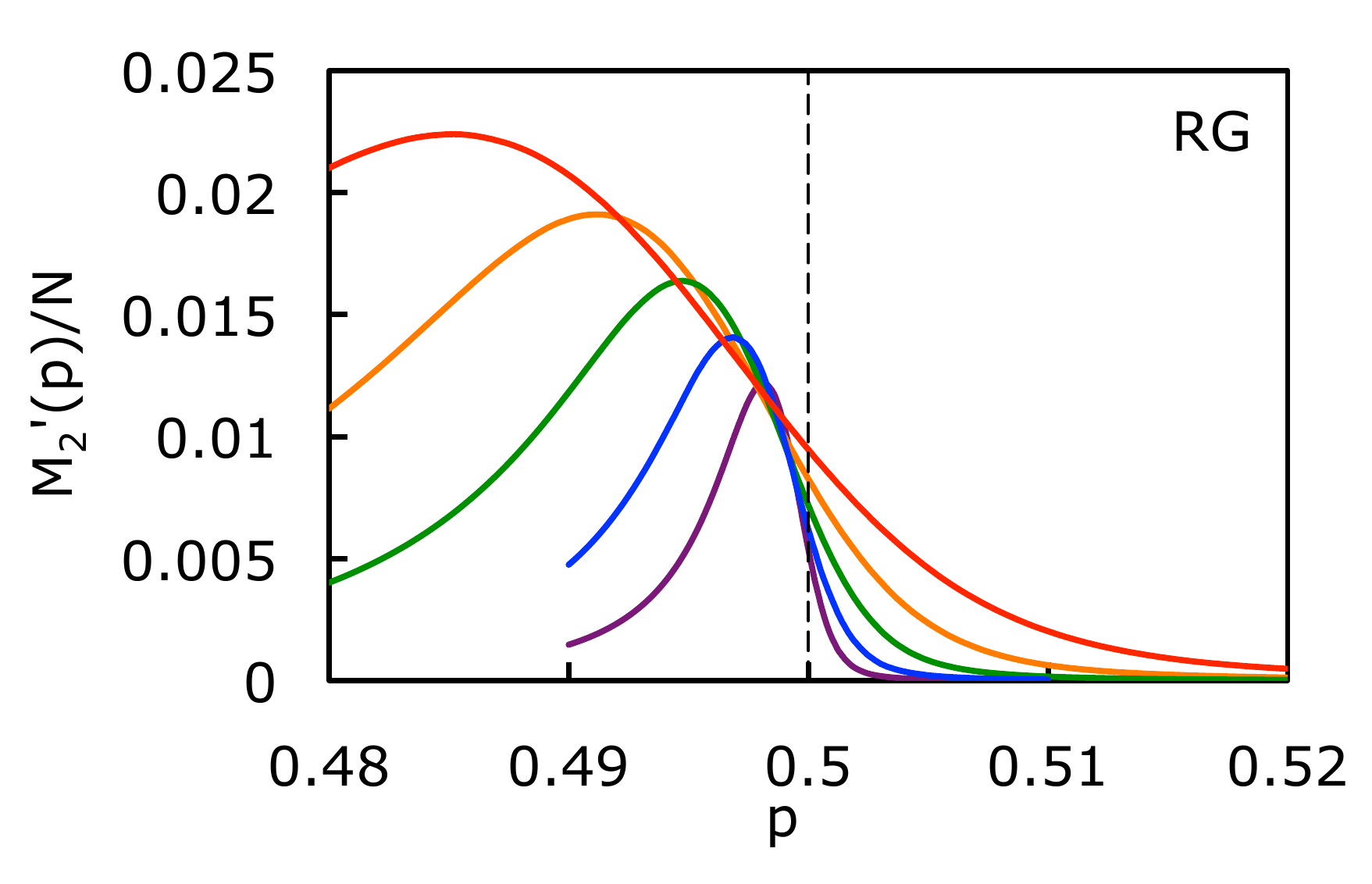} 
   \includegraphics[width=3in]{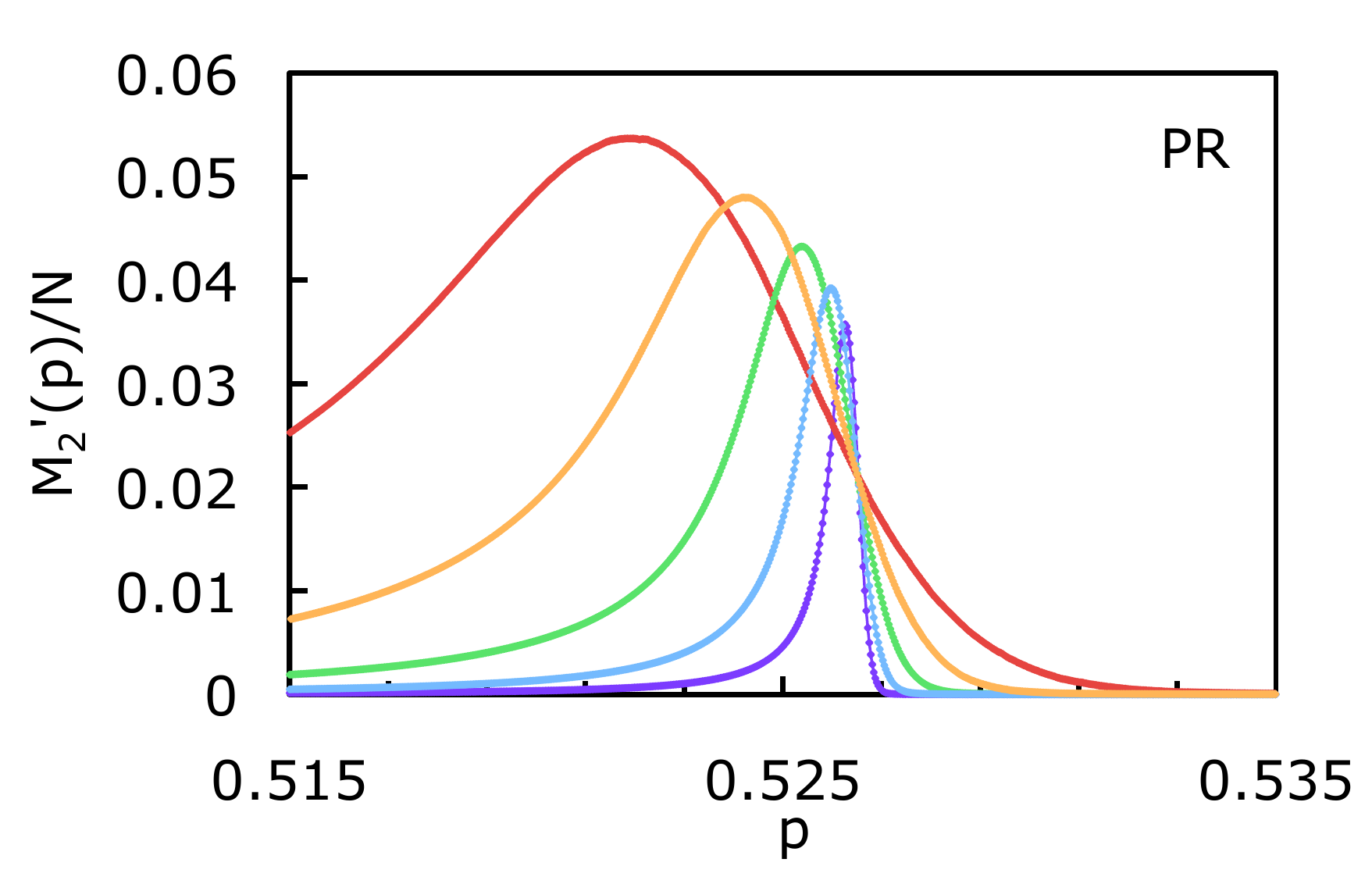} 
  \includegraphics[width=3in]{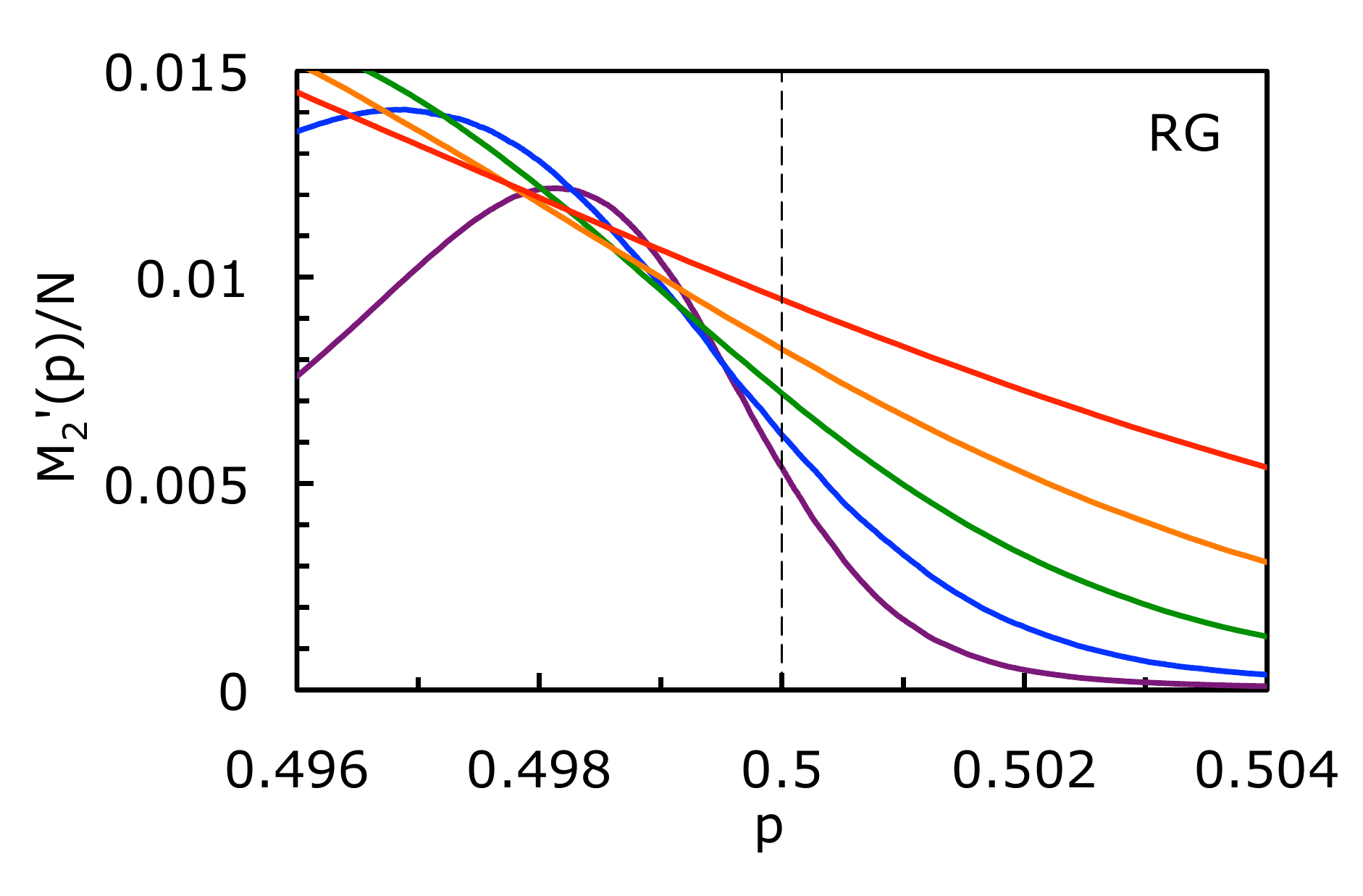} 
   \includegraphics[width=3in]{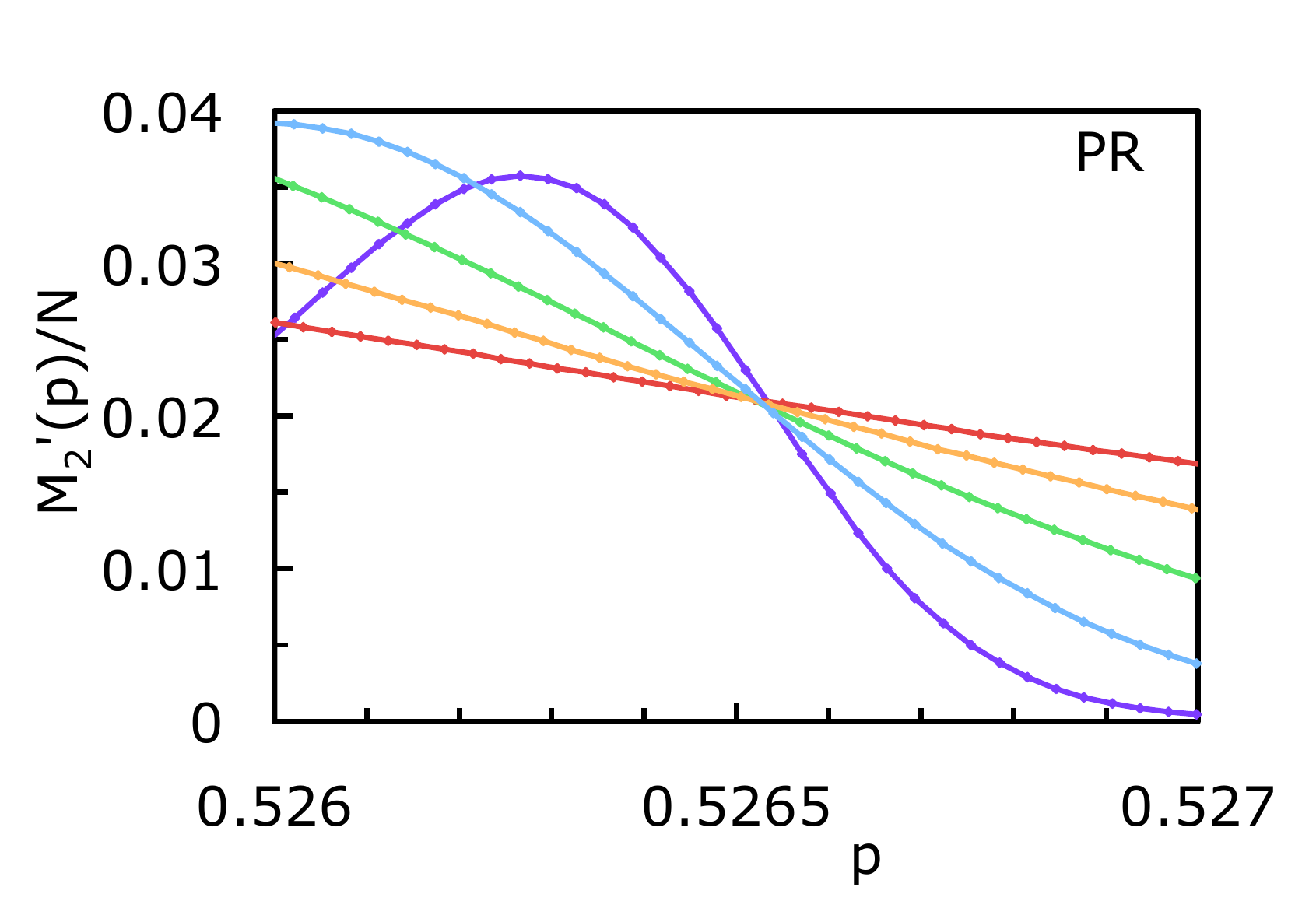} 
   \caption{(Color online) Scaled second moment minus largest cluster $M_2'(p)/N$, showing a distinct qualitative difference between the two models: the curves in the PR model cross at one point (presumably $p_c$), while those of the RG model do not.
   Lower plots show close-ups around $p_c$. }
   \label{fig:M2prime}
\end{figure*}

\section{Results}

The results of this work are shown in a series of pairs of figures, with results for the RG model 
placed on the left-hand side, and those for the PR model placed on the right-hand side.

\subsection{The maximum cluster size}

In Fig.\ \ref{fig:smax} we show the average of the maximum cluster size scaled by the number of sites,
 $\langle s_\mathrm{max} \rangle/N$, as a function
of $p$, for the different system sizes.  This quantity can also be identified with the usual order parameter, the percolation 
probability $P_\infty$, if one considers that the largest cluster is effectively the ``infinite" one.
The PR model (right panel) clearly shows qualitatively different behavior than the RG model, with crossing curves in
the PR case.

  \begin{figure*}[htbp] 
   \centering
   \includegraphics[width=3in]{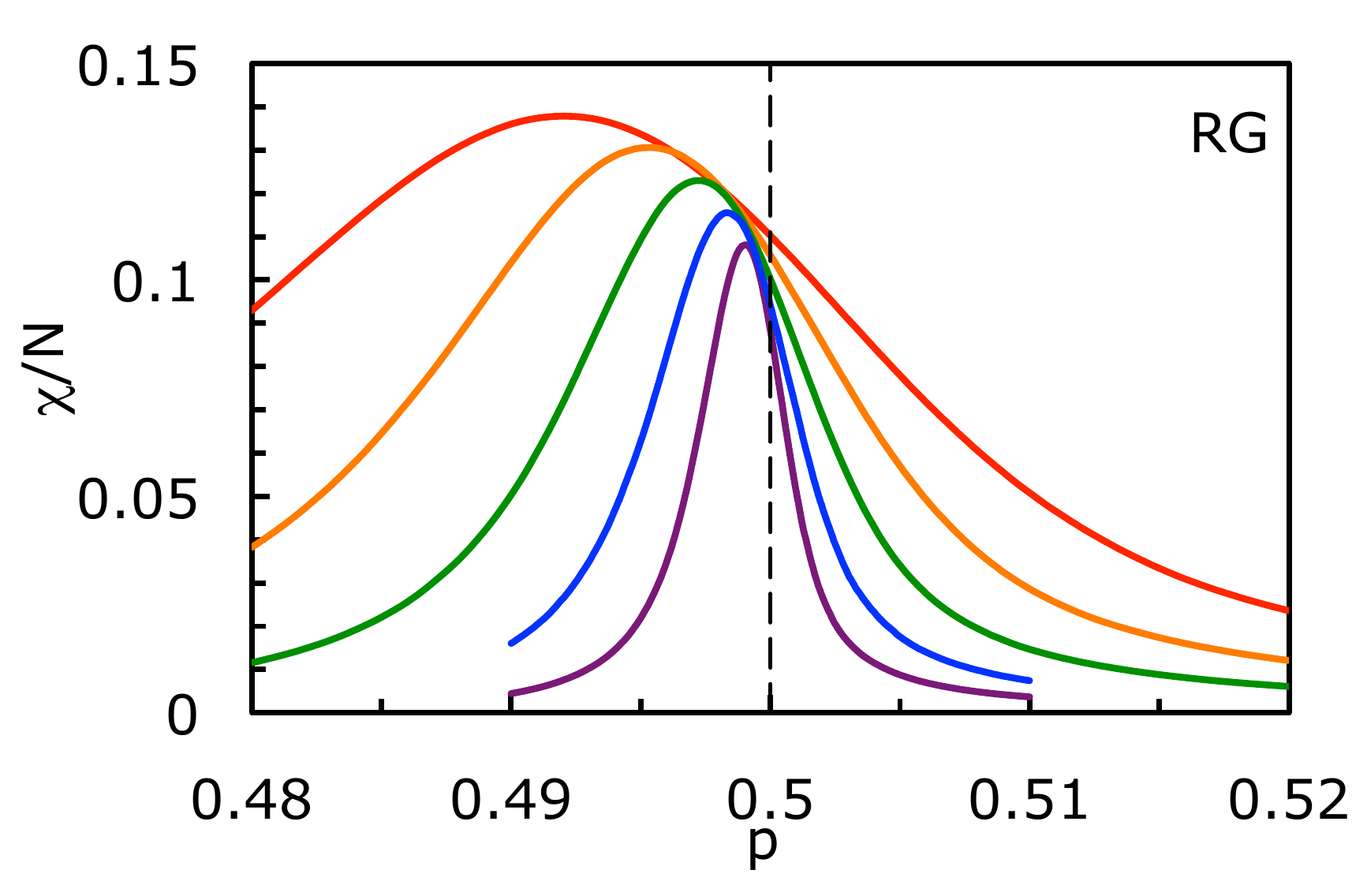} 
      \includegraphics[width=3in]{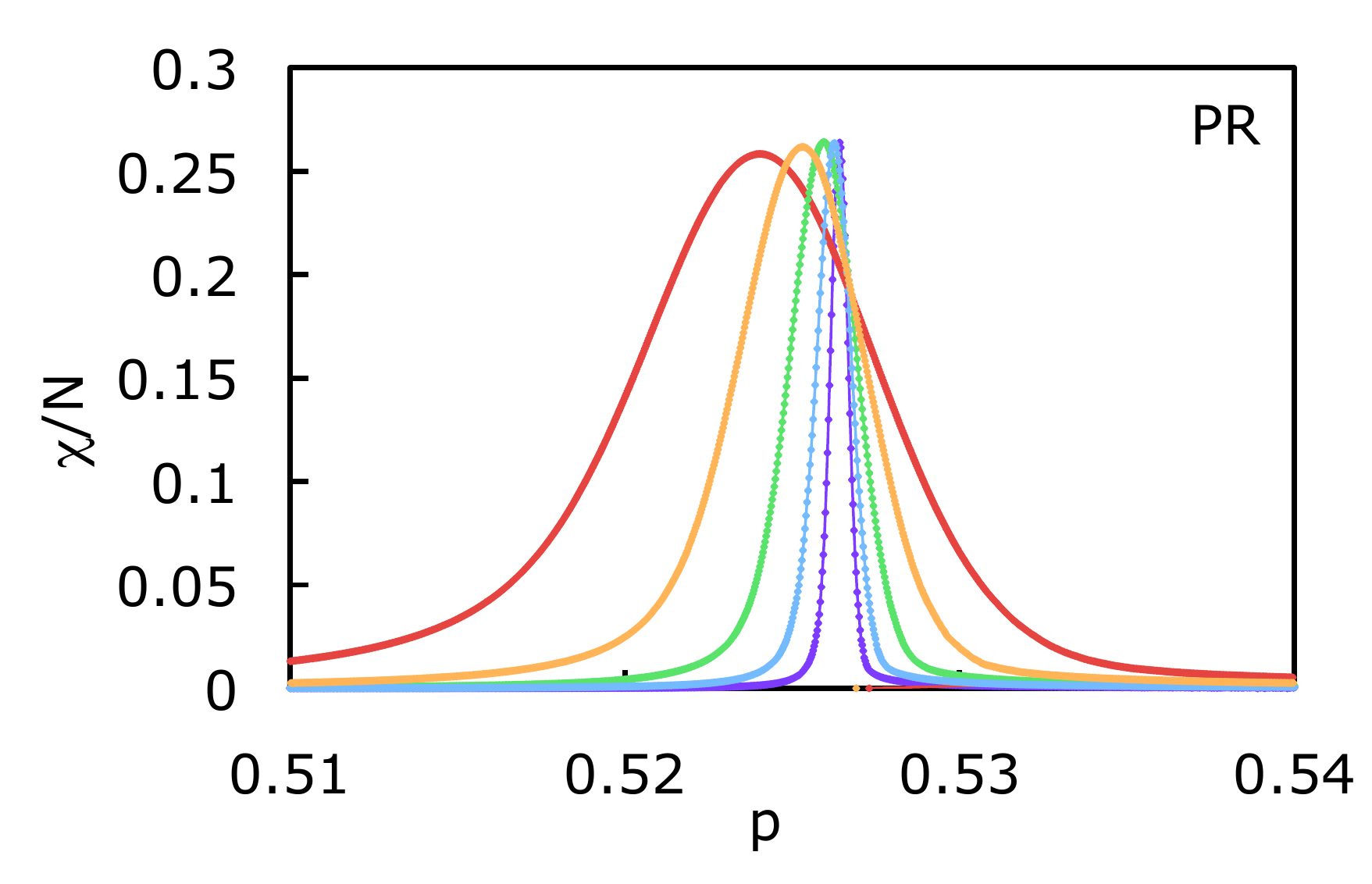} 
    \includegraphics[width=3in]{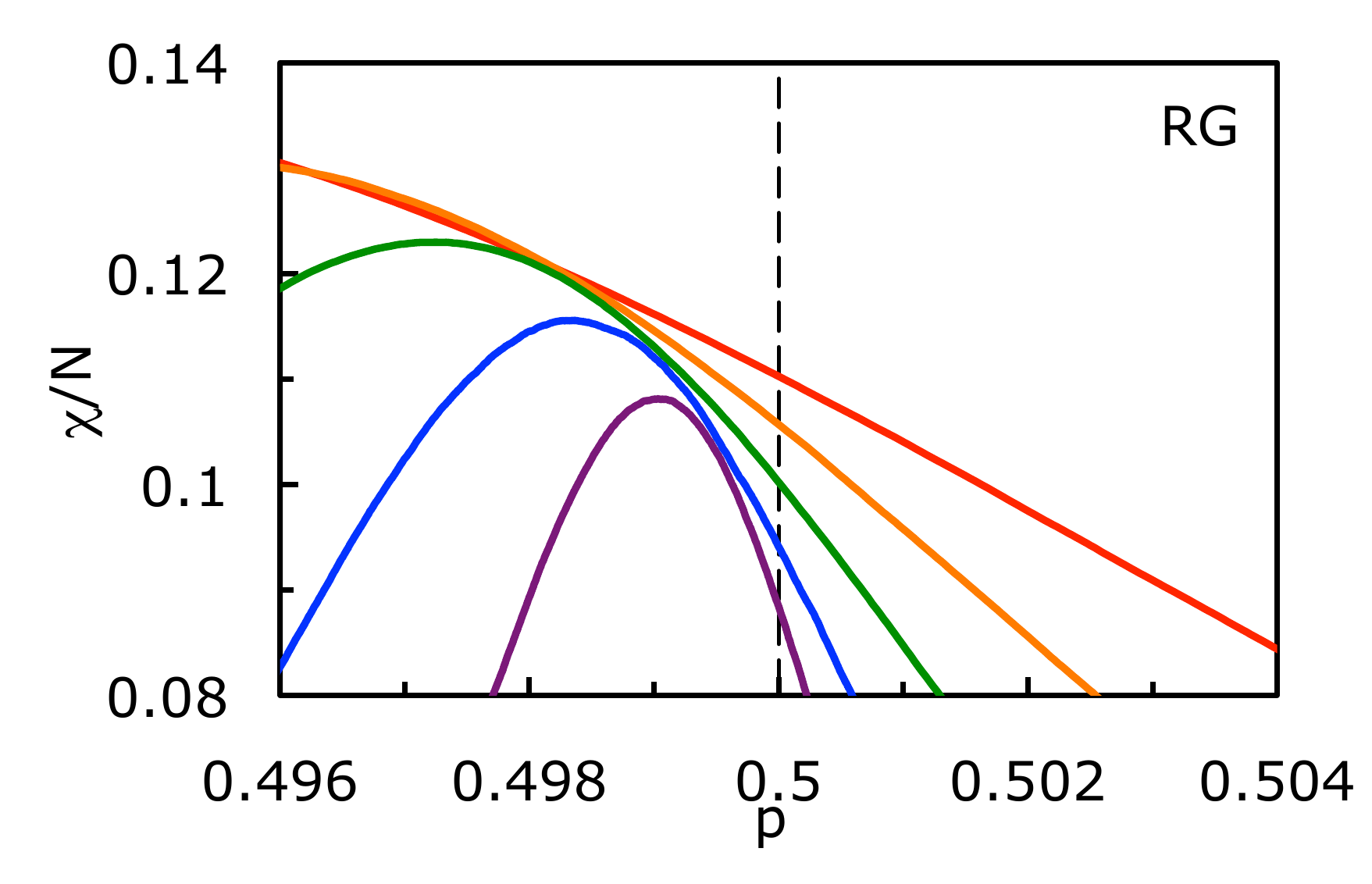} 
   \includegraphics[width=3in]{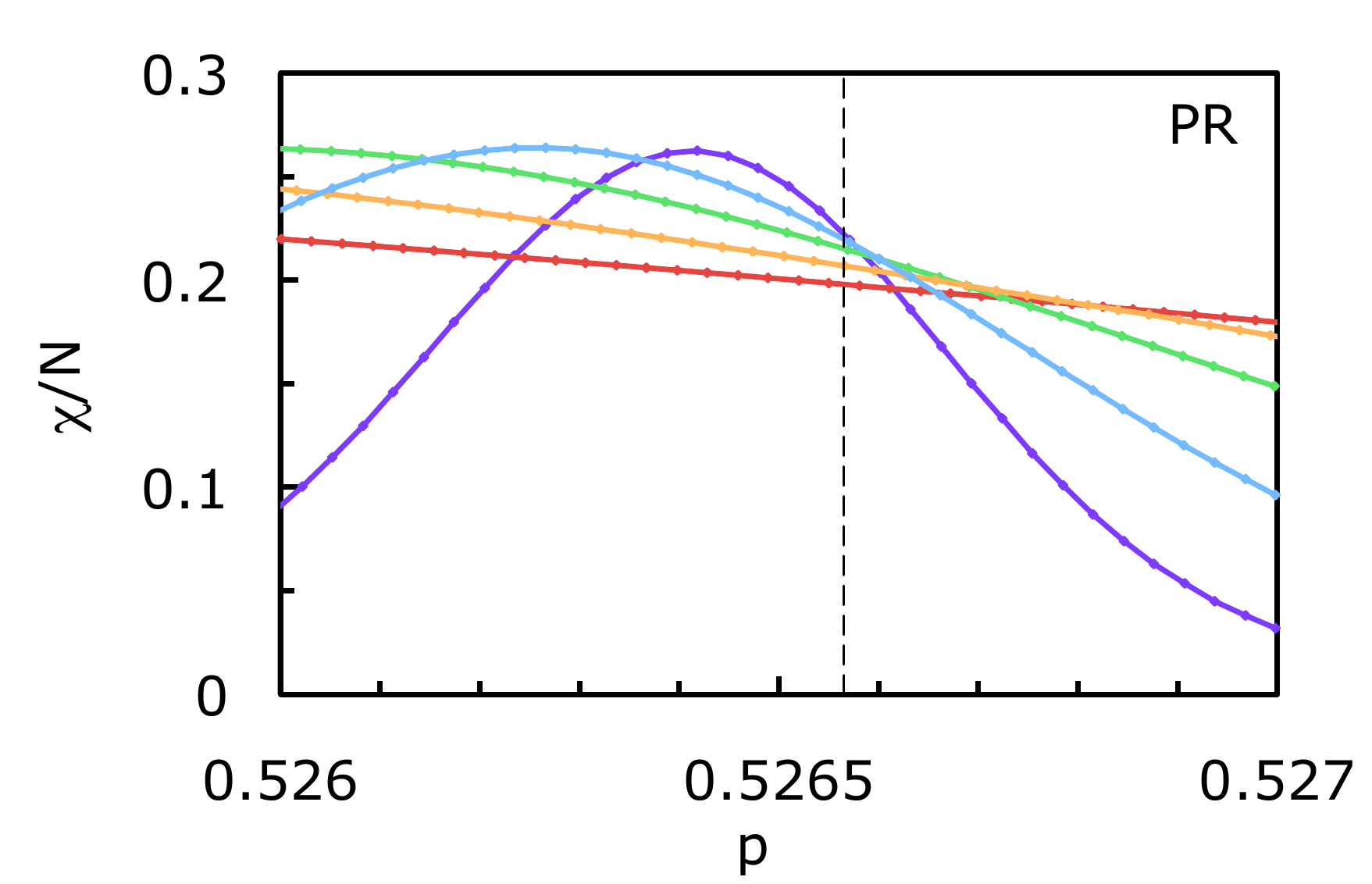} 
  \caption{(Color online) Susceptibility $\chi/N$ as a function of $p$.  Lower plots are close-ups of the
  behavior near $p_c$.}
   \label{fig:susceptibility}
\end{figure*}

\begin{figure*}[htbp] 
   \centering
   \includegraphics[width=3in]{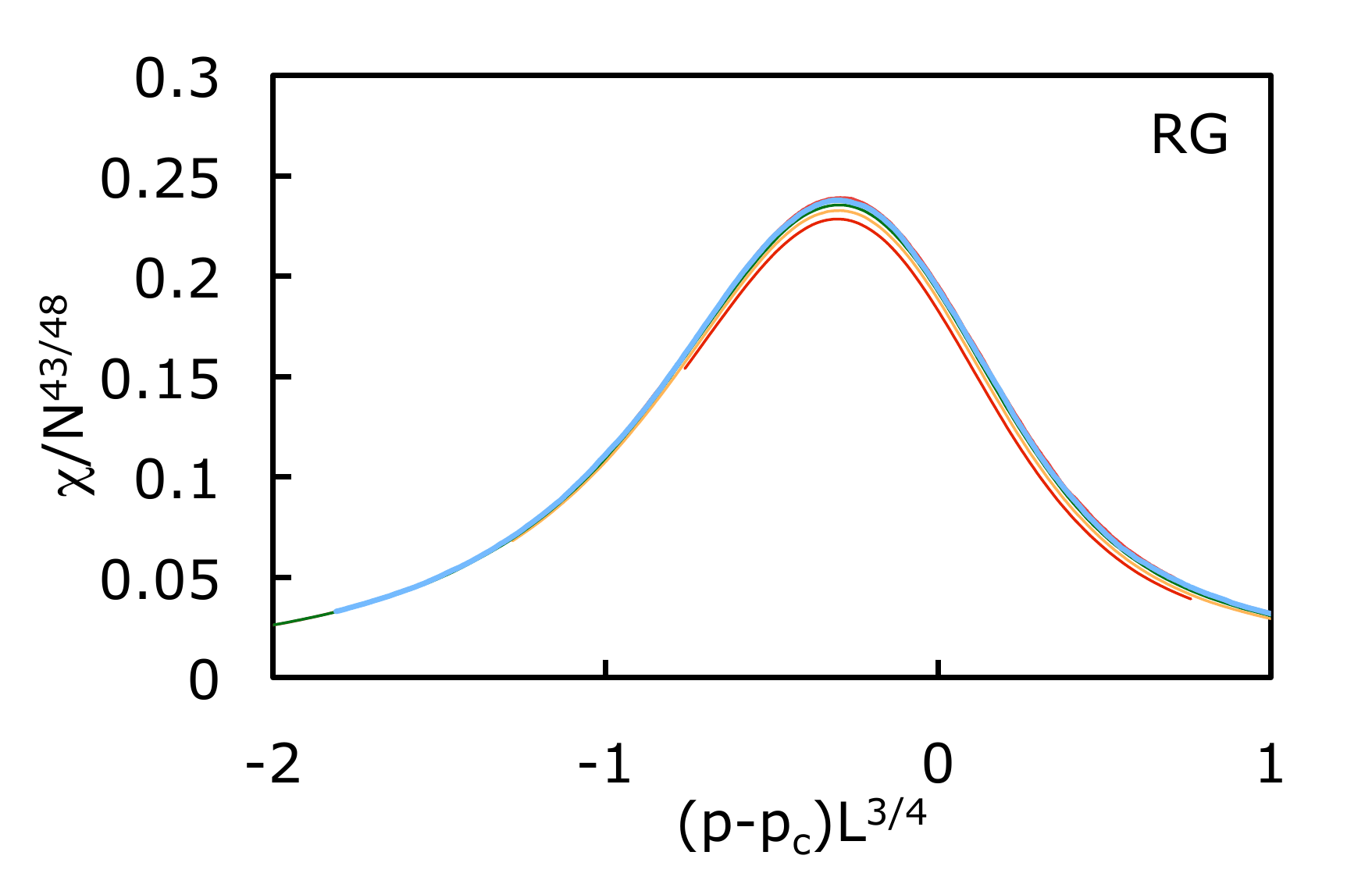} 
  \includegraphics[width=3in]{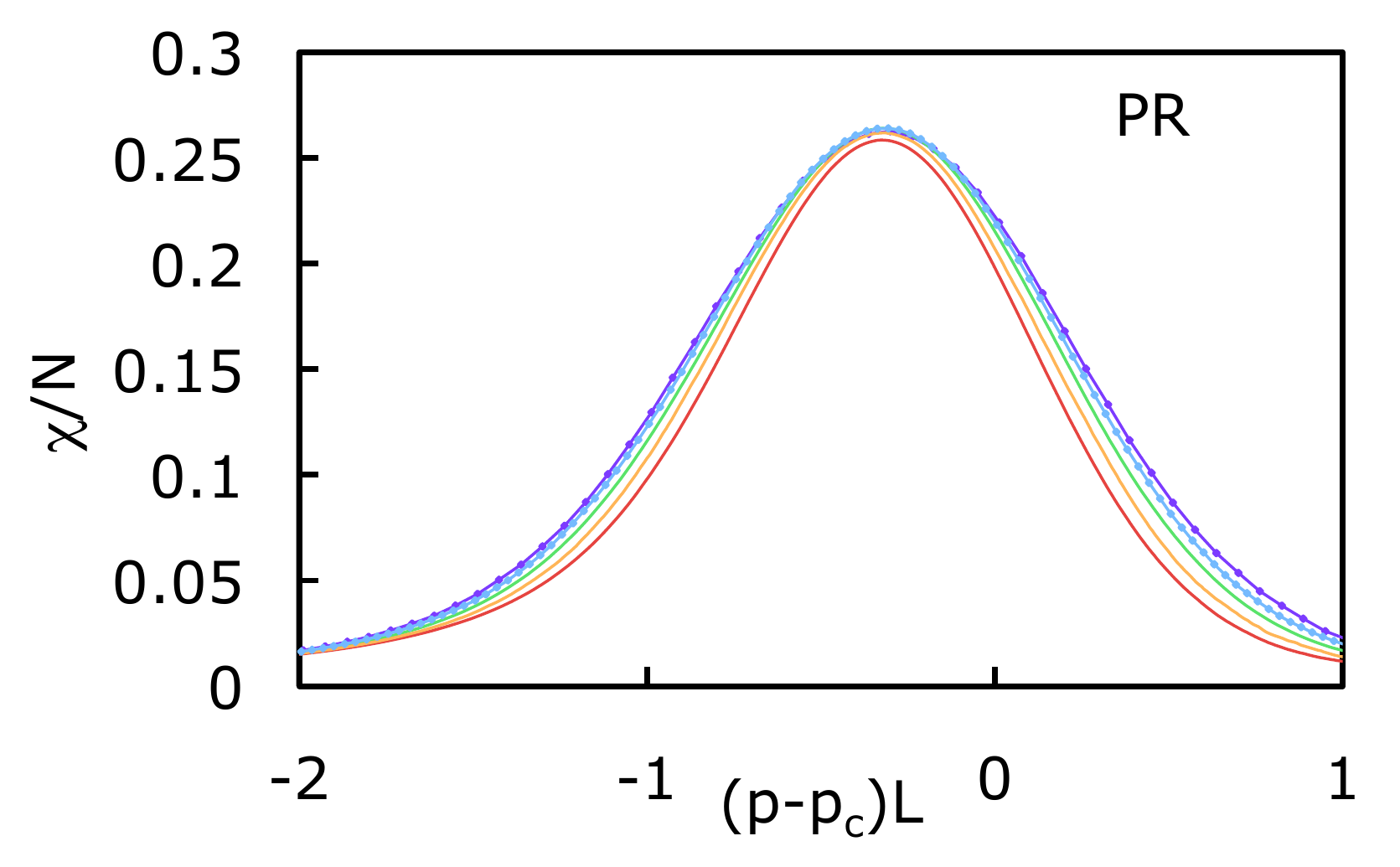} 
   \caption{(Color online) Scaling plots of the susceptibility,  assuming $\nu = 4/3$, $\gamma/2\nu = 43/48$ (RG), and $\nu = 1$, $\gamma/2\nu = 1$ (PR).  In both cases, curves $L = 128$ have the lowest peak, and $L = 2048$ have the highest peak.}
   \label{fig:susceptibilityscaling}
\end{figure*}

The behavior of $\langle s_\mathrm{max} \rangle /N$ at $p_c$ is shown in Fig.\ \ref{fig:M2smax}, using $p_c = 1/2$ for the
RG case and $p_c = 0.526565$ (determined below) for the PR case.
For the RG case, the slope ($-0.1062$) agrees within errors with the scaling predictions of $-\beta/\nu = -5/48 \approx- 0.104167$. 
The points for the PR model are also fit well by a straight line on the ln-ln plot, suggesting scaling for this quantity, with
slope $-\beta/\nu = -0.0589$, which is clearly different from the RG model.  Based upon the variation
with size, we estimate the error to this result to be $\pm 0.01$.
This value of $\beta/\nu$ is consistent with the value $\beta/\nu = 0.07(3)$ (within the error bars $\pm 0.03$) reported
in \cite{RadicchiFortunato10}.

\subsection{Moments and susceptibility}
Fig.\ \ref{fig:M2} shows the behavior of the second moment $M_2(p) = \sum_s s^2 n_s = (1/N) \sum_i s_i^2$, where $s_i$ is the
mass of the $i$-th particle, scaled by $N$.   Again, the PR model shows curve-crossing with a possible accumulation or crossing point.
The scaling behavior at $p_c$ is shown in Fig.\ \ref{fig:M2smax}.  The slope for the  RG model $-0.210$ is consistent  with the prediction $\gamma/\nu - 2 = -5/24 \approx -0.208333$. 
The PR data also appears to obey power-law behavior, with a slope $\gamma/\nu - 2 \approx -0.10$ implying
$\gamma/\nu \approx 1.90$, with an estimated error of 0.01.  This is
somewhat higher than the value $\gamma/\nu = 1.7(1)$ reported in \cite{RadicchiFortunato10}.

By scaling and  hyperscaling in 2d, one would expect that the slopes of the two curves in Fig.\ \ref{fig:M2smax} should differ by a factor of two:
 $\gamma/\nu - 2 = -2 \beta/\nu$.   This is seen to hold well for the RG data, but not so well for the PR case.
Further analysis of the data shows that the value $\gamma/\nu - 2 \approx -0.10$ seems to be independent of $L$, but 
$-\beta/\nu$ appears to be increasing as $L$ increases, and may possibly
approach the value $-0.05$ (as $L \to \infty$) implied by this scaling.  However, studies on larger systems are
needed to confirm this.

\begin{figure*}[htbp] 
   \centering
   \includegraphics[width=3in]{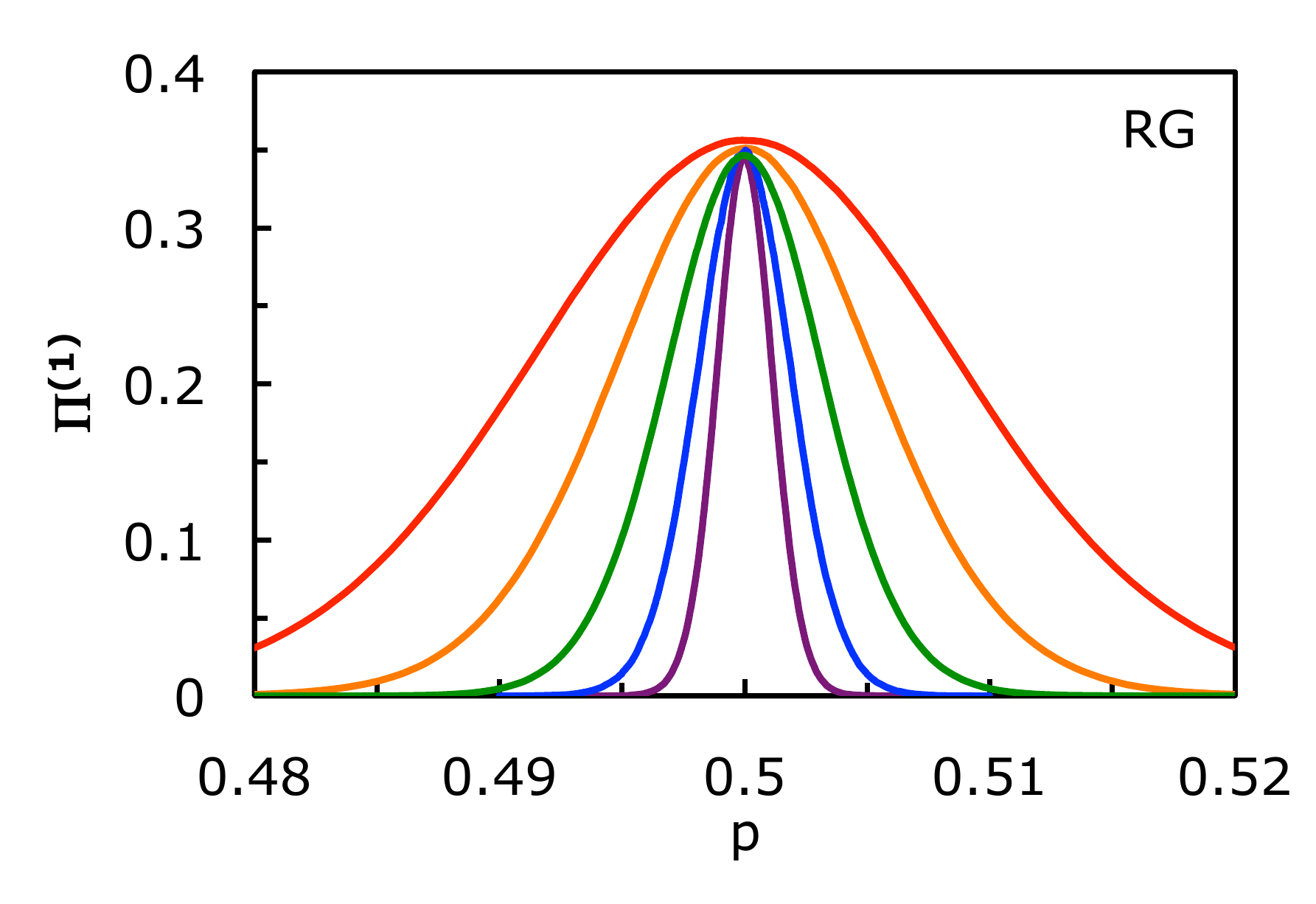} 
      \includegraphics[width=3in]{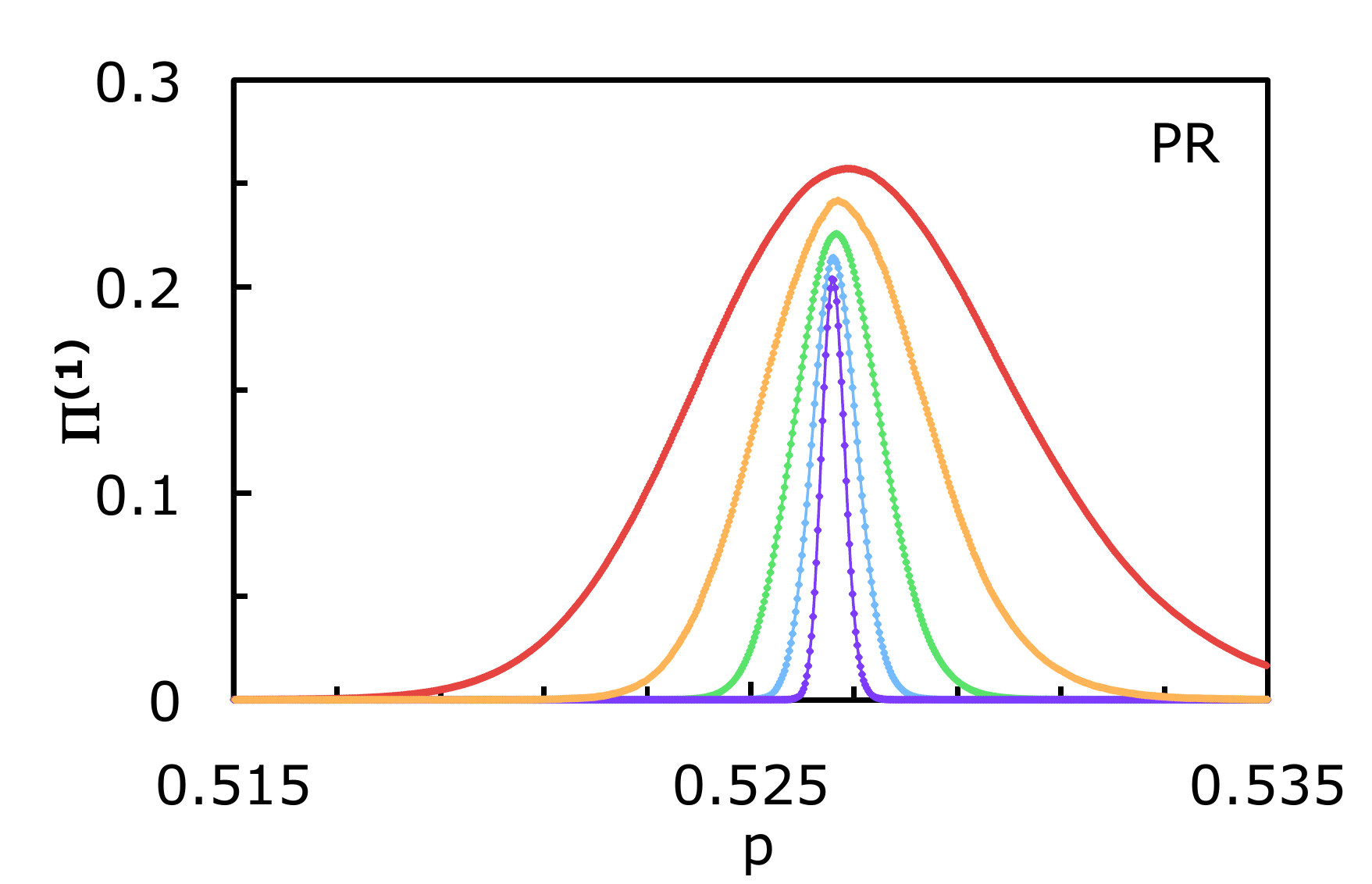} 
   \caption{(Color online) One-way wrapping probability $\Pi^{(1)}$ as a function of $p$. The width of the distribution is plotted in 
   Fig.\ \ref{fig:onewaywidth} and the mean $p$ values are plotted in Fig.\ \ref{fig:pcestimates}.}
   \label{fig:oneway}
\end{figure*}

\begin{figure*}[htbp] 
   \centering
   \includegraphics[width=3in]{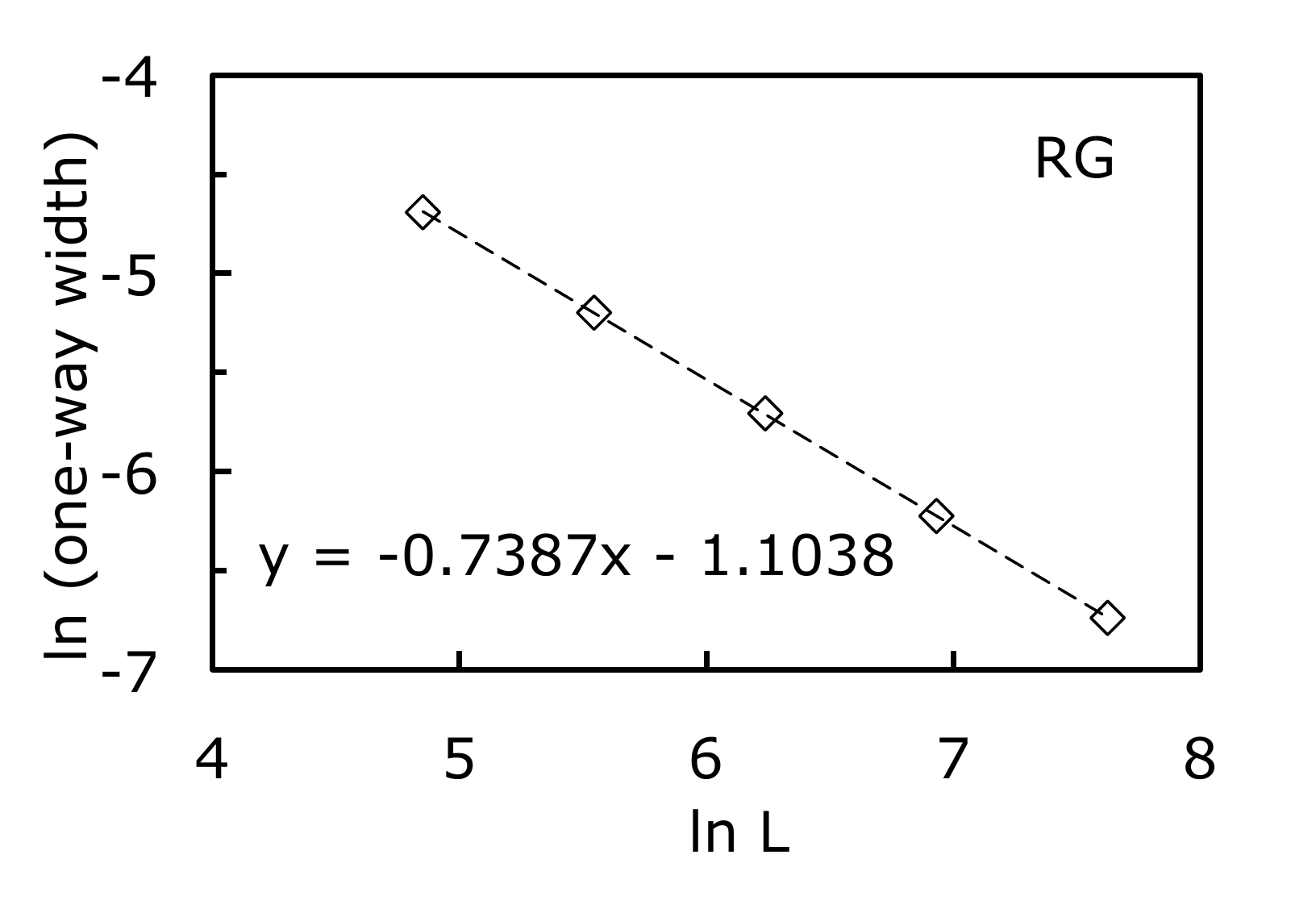} 
      \includegraphics[width=3in]{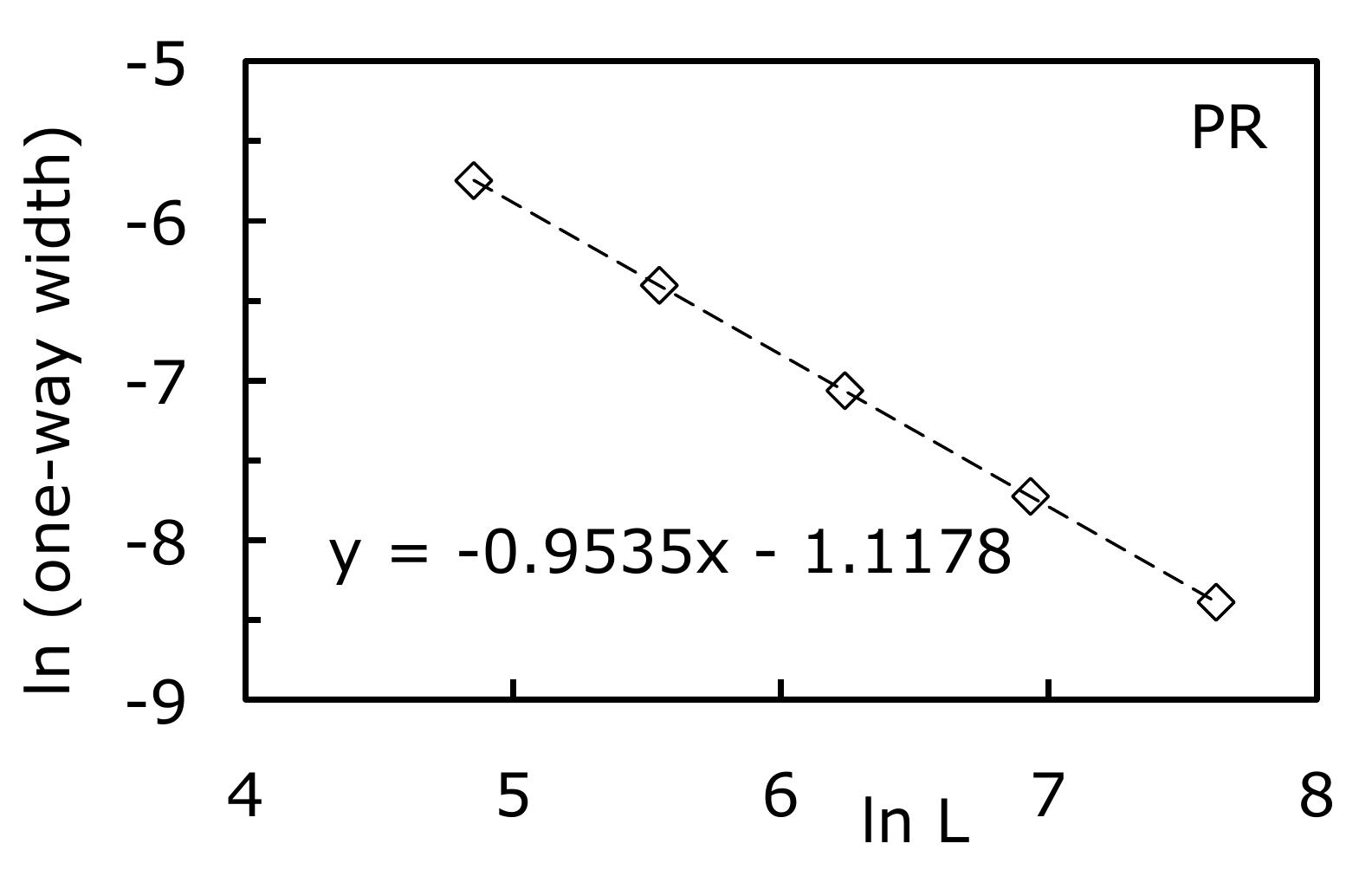} 
   \caption{The $\ln$ of the width of $\Pi^{(1)}(p)$ (shown in Fig.\ \ref{fig:oneway}) as a function of $\ln L$.
   The linear fit to the points is shown on the plot, where $y$ represents the $\ln$ of the width and $x$ represents $\ln L$.  The slope is consistent with $1/\nu = 3/4$ for the RG case, and suggests $1/\nu \approx 0.95$ in the PR case.}
   \label{fig:onewaywidth}
\end{figure*}

In Fig.\ \ref{fig:M2prime} we show the behavior of the scaled second moment minus the largest cluster, that is:
\begin{equation}
\frac{M_2'}{N} = \frac{1}{N^2} \sum_{i \ne \mathrm{max}} s_i^2 = \frac{M_2}{N} - \frac{\langle s_\mathrm{max}^2 \ .\rangle}{N^2}
\end{equation} 
According to scaling arguments, this function should go through a maximum at a value $p = p_\mathrm{max}$ where the  $(p_\mathrm{max}-p_c)L^{1/\nu}$ is a certain constant, at which point, $M_2'(p_\mathrm{max})/N$ should scale as $L^{\gamma/\nu-2}$.  We verified that the
peaks for RG in Fig.\ \ref{fig:M2prime} obey this behavior with standard exponents (not shown).  However, for the PR model, the curves of $M_2'/N$ very closely pivot
 around the crossing point at $p_c \approx 0.52654$, which is also close to
 the inflection points of the curves.  This suggests that as $L \to \infty$, $M_2'(p_c)/N$ is non-zero, 
 which would imply that $\gamma/\nu = 2$, in conflict with what we found ($\gamma/\nu \approx 1.90$) from the behavior of 
 $M_2 (p_c)$.  This behavior is another indication of the unusual nature of the PR transition.

In Fig.\ \ref{fig:susceptibility} we show the behavior of the susceptibility $\chi$, defined by
\begin{equation}
\chi = \sqrt{\langle s_\mathrm{max}^2\rangle  - \langle s_\mathrm{max}\rangle^2} \ ,
\end{equation}
which characterizes the fluctuations in the size of the largest cluster.  It can also be found 
from previous quantities via
$\chi = (M_2 - M_2' - \langle s_\mathrm{max} \rangle^2/N)^{1/2}$.
The peaks of $\chi(p_c)/N$ in the RG model decay to zero as $L^{-0.22}$, consistent with the scaling prediction $L^{\gamma/\nu-2} = L^{-5/24}$.  However, the peaks in the PR model apparently increase to a constant value  $\chi/N \approx 0.264$, again consistent with $\gamma/\nu = 2$.
Also, the locations of $p$ at the peaks for the PR model approach $p_c \approx 0.526575$ as $L^{-1}$,again  implying $\nu = 1$.  
Below
we will find that several other quantities also satisfy inverse-size scaling (Fig.\ \ref{fig:pcestimates}). 

In Fig.\ \ref{fig:susceptibilityscaling}, we show a scaling plot of $\xi/N$ vs.\ $(p-p_c)L$ assuming $\nu = 1$, and also
$\gamma/\nu = 2$, and the fit is seen to be good.  (Taking $\nu = 1/0.96$ yields a much poorer fit.)  A similar plot for the RG model, with standard percolation
scaling, is shown for comparison.


\begin{figure*}[htbp] 
   \centering
   \includegraphics[width=3in]{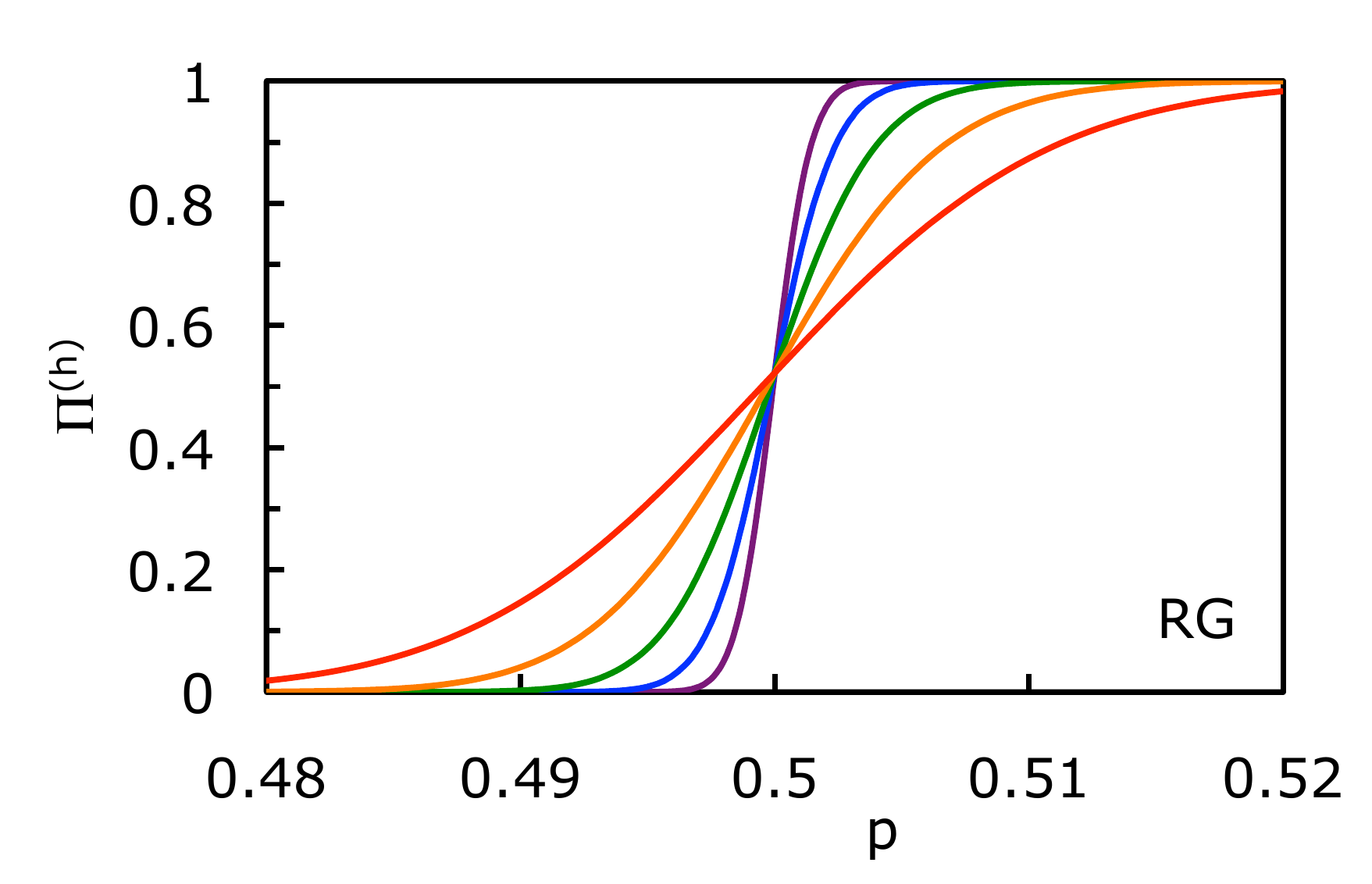} 
      \includegraphics[width=3in]{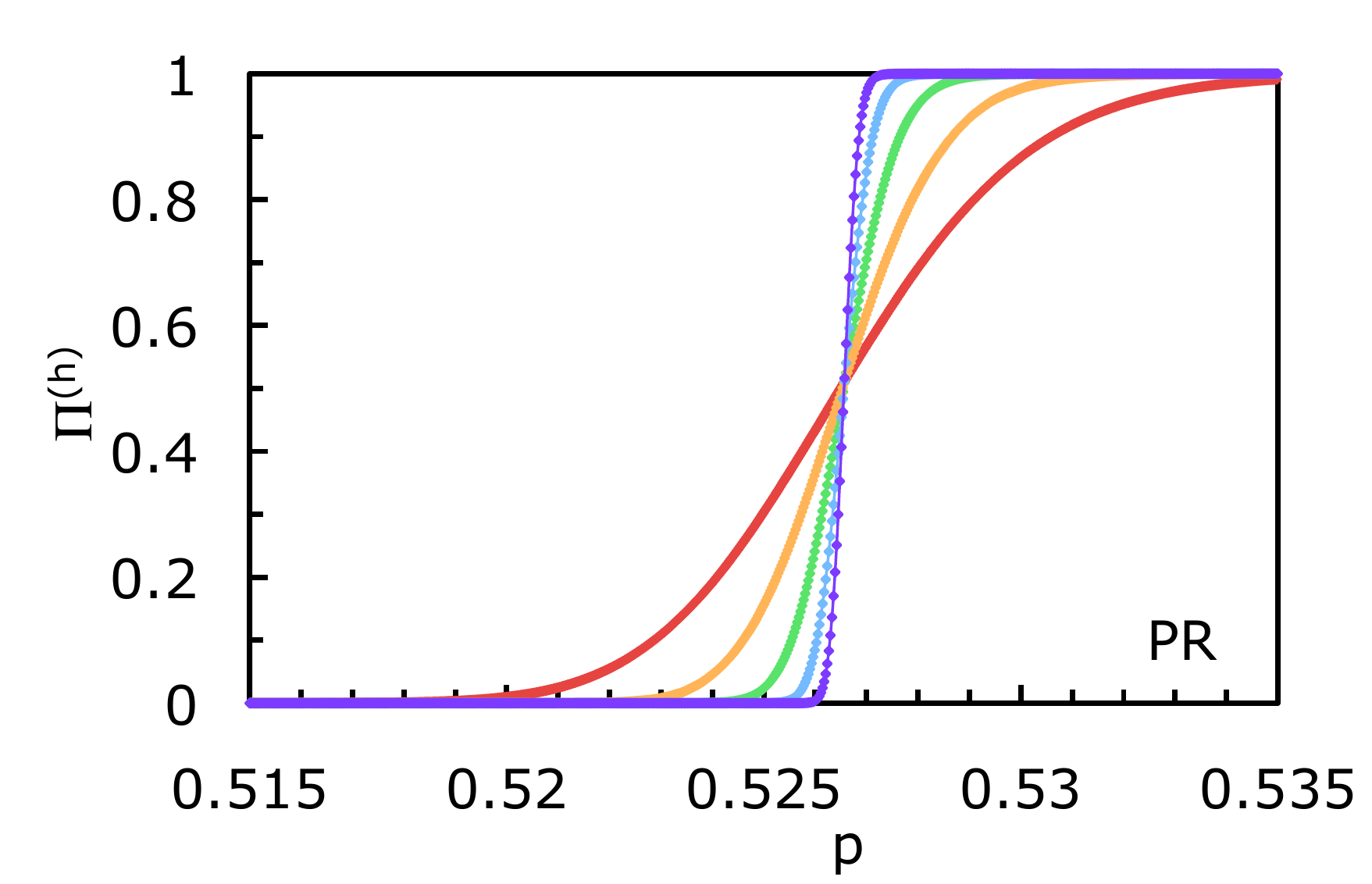} 
   \includegraphics[width=3in]{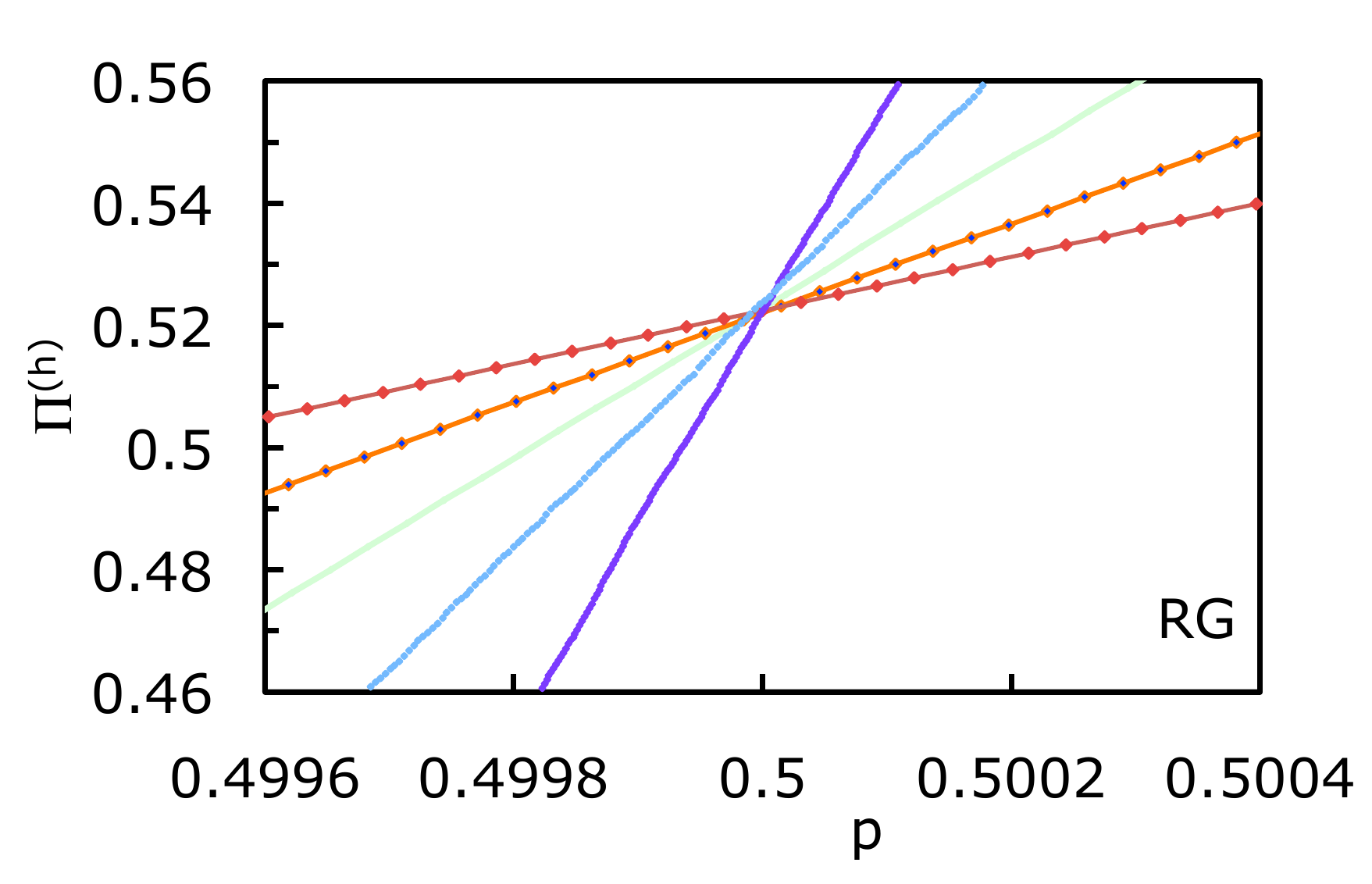} 
      \includegraphics[width=3in]{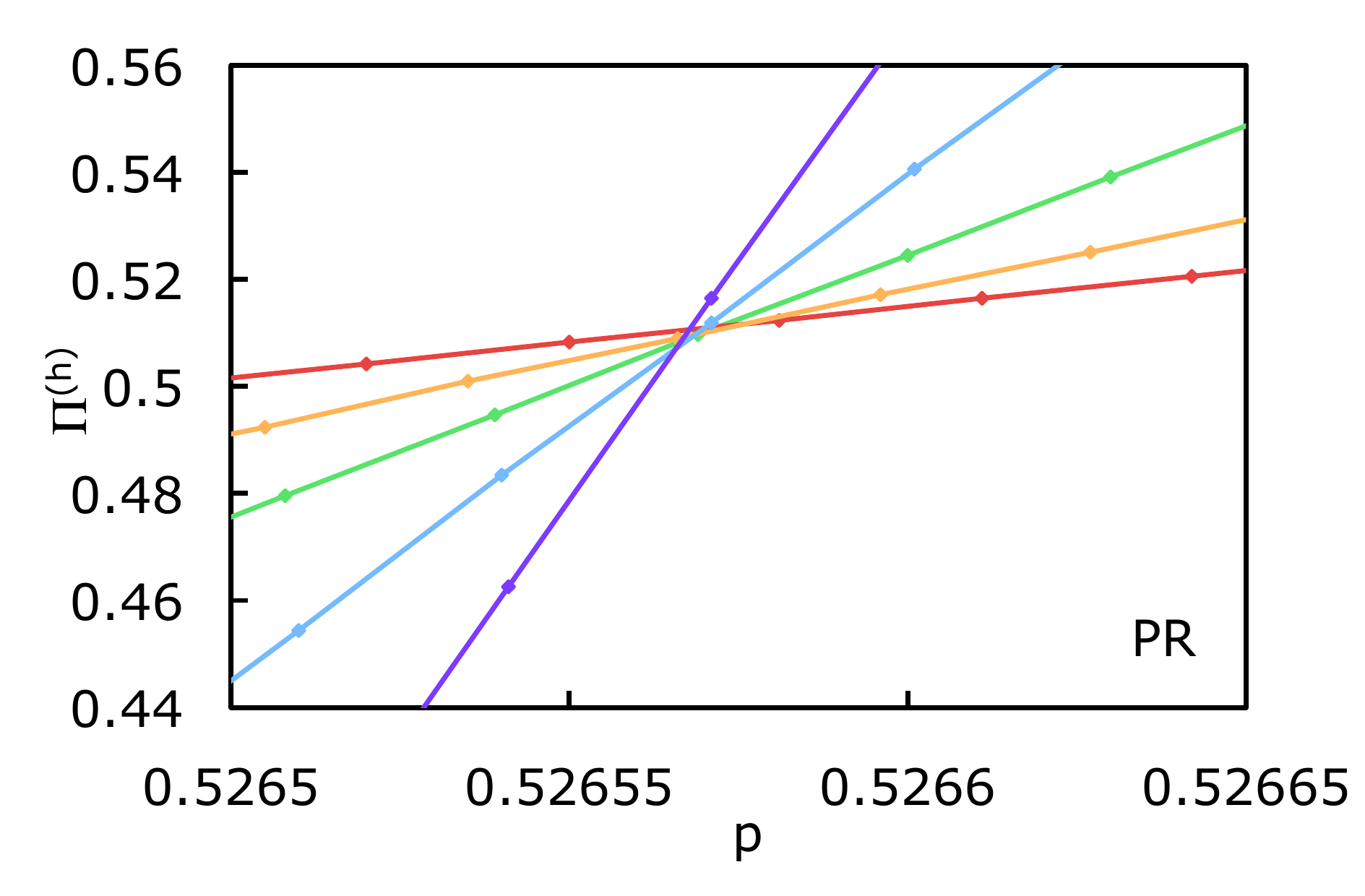} 
   \caption{(Color online) Horizontal wrapping probability.  Expanded plots are given in the
   lower panels, which show the precise crossing of the curves for different $L$; note that the horizontal scale for the PR model is more
   expanded than that of the RG model.}
   \label{fig:horiz}
\end{figure*}

\subsection{Wrapping probabilities}

Next we consider wrapping probabilities.  For standard percolation these were studied theoretically by Pinson \cite{Pinson94} and numerically in \cite{ZiffLorenzKleban99,NewmanZiff01,OliveiraNobregaStauffer03}.   This work has also been generalized to the Potts model \cite{Arguin02,MorinDuchesneSaintAubin09}.   Even though the percolating critical cluster is a fractal and of zero density in the continuum limit, the wrapping probability remains finite and has a value that depends only upon the aspect ratio of the system and the type of wrapping homology.

In Fig.\ \ref{fig:oneway}, we show the (only) one-way wrapping probability $\Pi^{(1)}$, defined as 
the probability at least one cluster wraps horizontally but not vertically, or wraps vertically but not horizontally.  For the RG model, the value of $\Pi^{(1)}$ at $p_c = 1/2$ approaches the predicted value $0.351642855\ldots$ \cite{Pinson94,NewmanZiff01} very rapidly.  The curves are exactly symmetric, because for one-way wrapping there must also be a one-way wrapping on the dual lattice, which in the square system is identical to the original lattice but with bonds occupied with probability $1 - p$.  For the PR model, the curves are not quite symmetric, and the value of $p$ at the peaks approaches $0.52658$ apparently as $L^{-1}$ (not shown).  The value of $\Pi^{(1)}$  seems to be dropping to a constant value of about $0.18$ as $\approx L^{-0.5}$, although the range of values of $L$ we considered is not sufficient to be very certain about this behavior.

The width (standard deviation) of $\Pi^{(1)}(p)$, as a function of $L$, is shown in Fig.\ \ref{fig:onewaywidth}.  For RG, the data are consistent with 
the theoretical prediction of a straight line with slope of $-1/\nu = -0.75$.  For the PR, the overall slope of the points is $-0.95$ but decreases to $-0.96$ for
large $L$, implying that $\nu \approx 1/0.96$.  This is in contrast with the value $\nu \approx 1$ seen in several other situations.

In Fig.\ \ref{fig:horiz} we show the probability distribution $\Pi^{(h)}$ for horizontal wrapping, irrespective of whether wrapping occurs in the vertical direction.  For both the RG and PR models, the curves cross at a single point, within numerical error.  
The crossing point of the RG model is at  $p = 0.499995(5)$, $\Pi^{(h)} = 0.5210$,  consistent with Pinson's theoretical result  $\Pi^{(h)}(p_c) = 0.52105829\ldots$ 
\cite{Pinson94,NewmanZiff01}, while that for the PR model is at $p = 0.526566(3)$, $\Pi^{(h)} = 0.5106$.  
Convergence behavior of the ordinary percolation crossing point was studied in \cite{NewmanZiff01}, however for site percolation and in the grand-canonical (fixed $p$) rather than the canonical (fixed-$n$)
ensemble.  We have not determined the convergence 
in this case, but the crossing point for the system sizes we considered clearly gives a very precise indication of $p_c$.

\begin{figure*}[htbp] 
   \centering
   \includegraphics[width=3in]{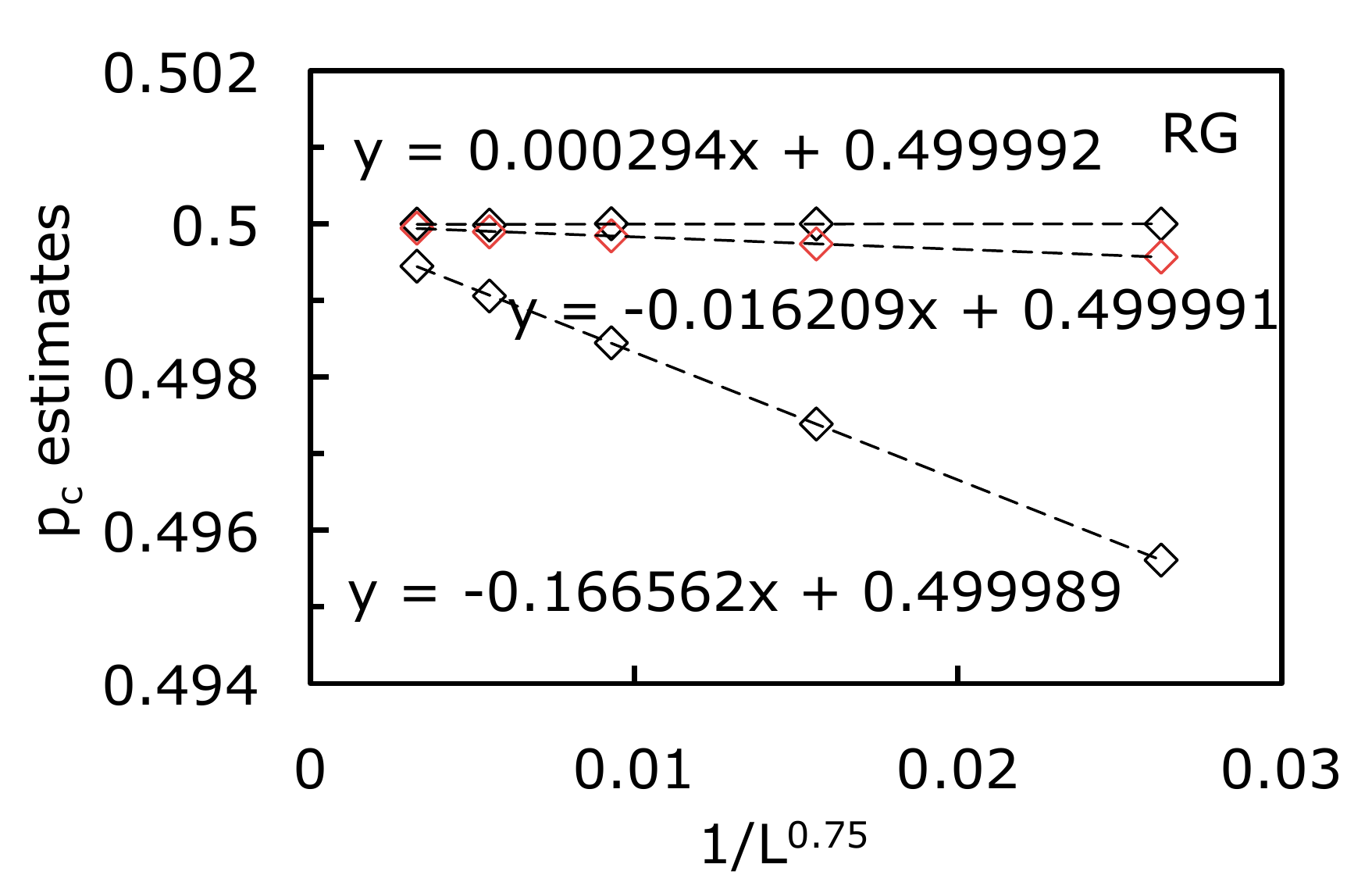} 
   \includegraphics[width=3in]{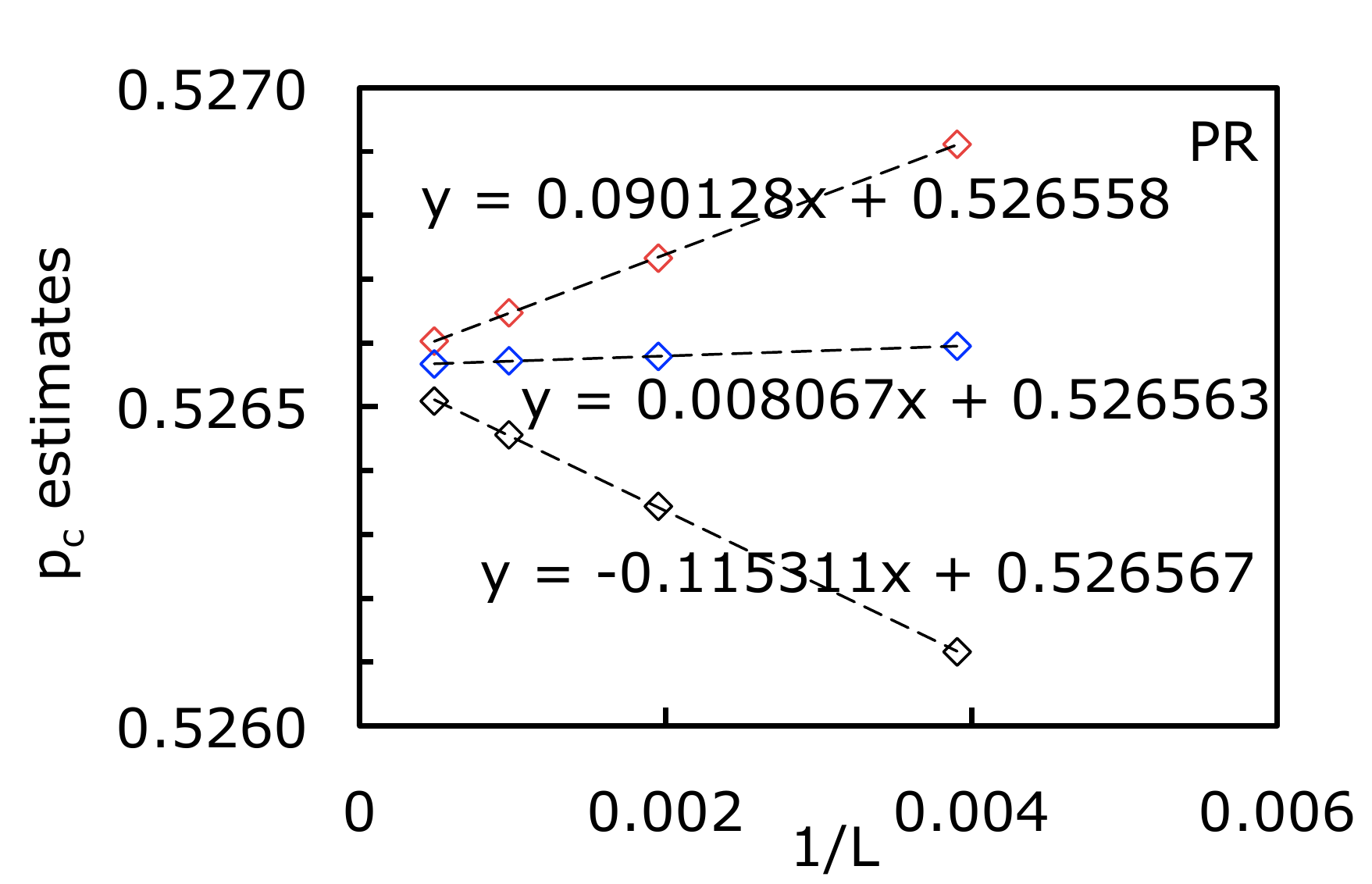} 
   \caption{(Color online) Estimates of $p_c$ vs.\ $L^{-0.75}$ (RG) and $L^{-1}$ (PR). In each case, we have 
   estimates determined from the average of $p$ at which one-way wrapping occurs  (top), the average
   value of $p$ 
   at which horizontal wrapping \emph{first} occurs (middle), and the average value of $p$ at which  either horizontal or vertical 
   crossing \emph{first} occurs (bottom).  Linear fits to the data are shown in the figure, where $y$ represents the $p_c$ estimates and $x$ represents $1/L$. }
   \label{fig:pcestimates}
\end{figure*}

The estimates for $p_c$ that come from various measures are summarized in Fig.\ \ref{fig:pcestimates}, plotted as a function of  $L^{1/\nu} = L^{-0.75}$ (RG) and $L^{-1}$ (PR).  The upper curves show the average of $p$ at which  one-way wrapping occurs --- that is, the mean  of the distribution shown in  
Fig.\ \ref{fig:oneway},  $\sum p \Pi^{(1)}(p)$.    The middle curves show the average value of $p$ 
   at which horizontal wrapping first occurs; for the PR case, this is nearly horizontal, so this quantity
is very good for estimating $p_c$ precisely.  The lower curves show 
   the average value of $p$ at which \emph{either} horizontal \emph{or} vertical  
   crossing first occurs.   For the RG model, all estimates extrapolate to a value very close to the expected value $0.5$, and for the PR 
   model the extrapolations are consistent with $p_c = 0.526263(3)$.  
   
  Note that here, we find a better fit to the data assuming $\nu = 1$ rather than $\nu \approx 1/0.96$ found in the scaling
  of the one-way width, Fig.\ \ref{fig:onewaywidth}.   However, 
  if we use the latter value, we don't find a significant change in the estimated value of $p_c$.

   \begin{figure*}[htbp] 
   \centering
   \includegraphics[width=3in]{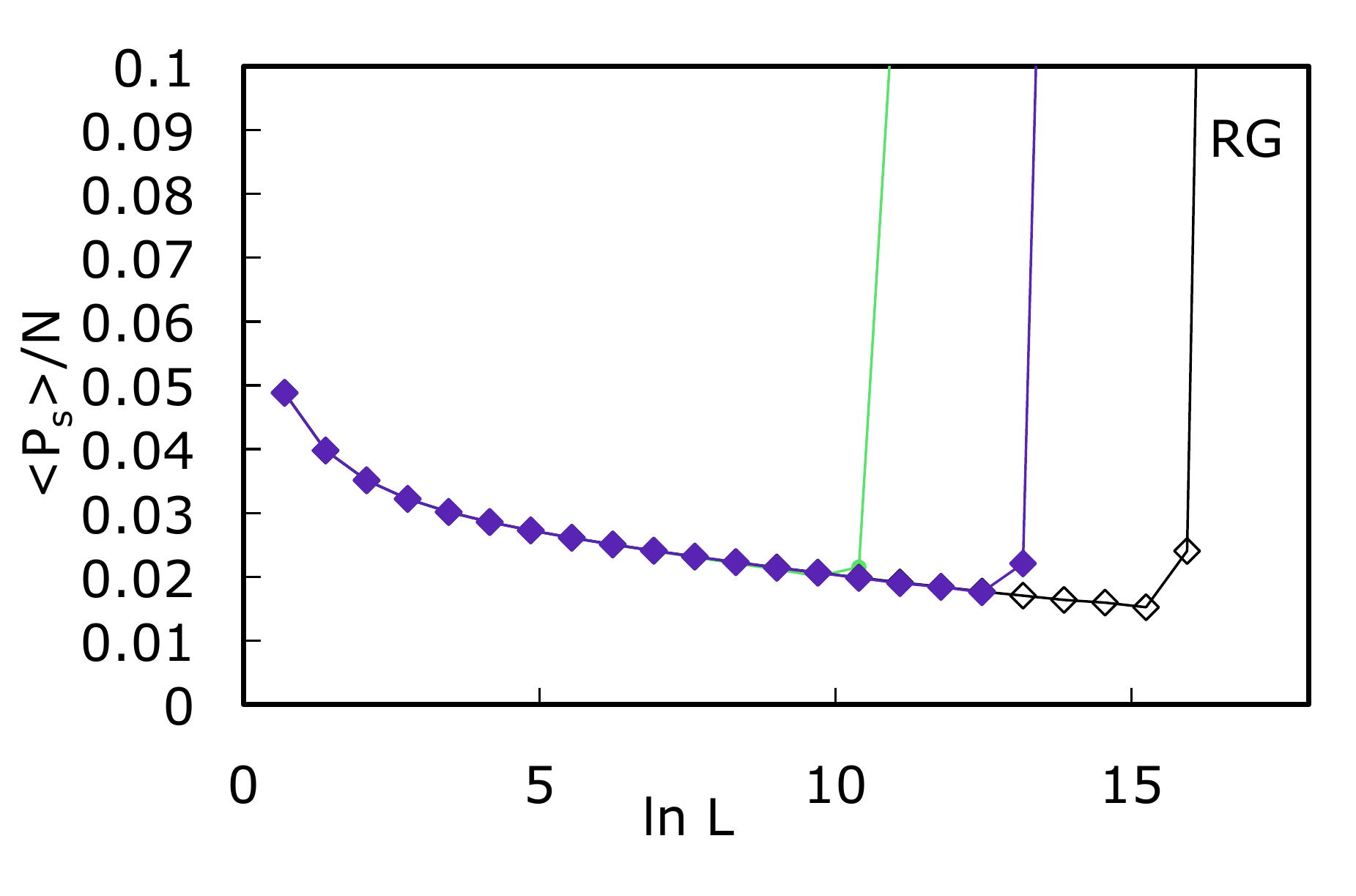} 
   \includegraphics[width=3in]{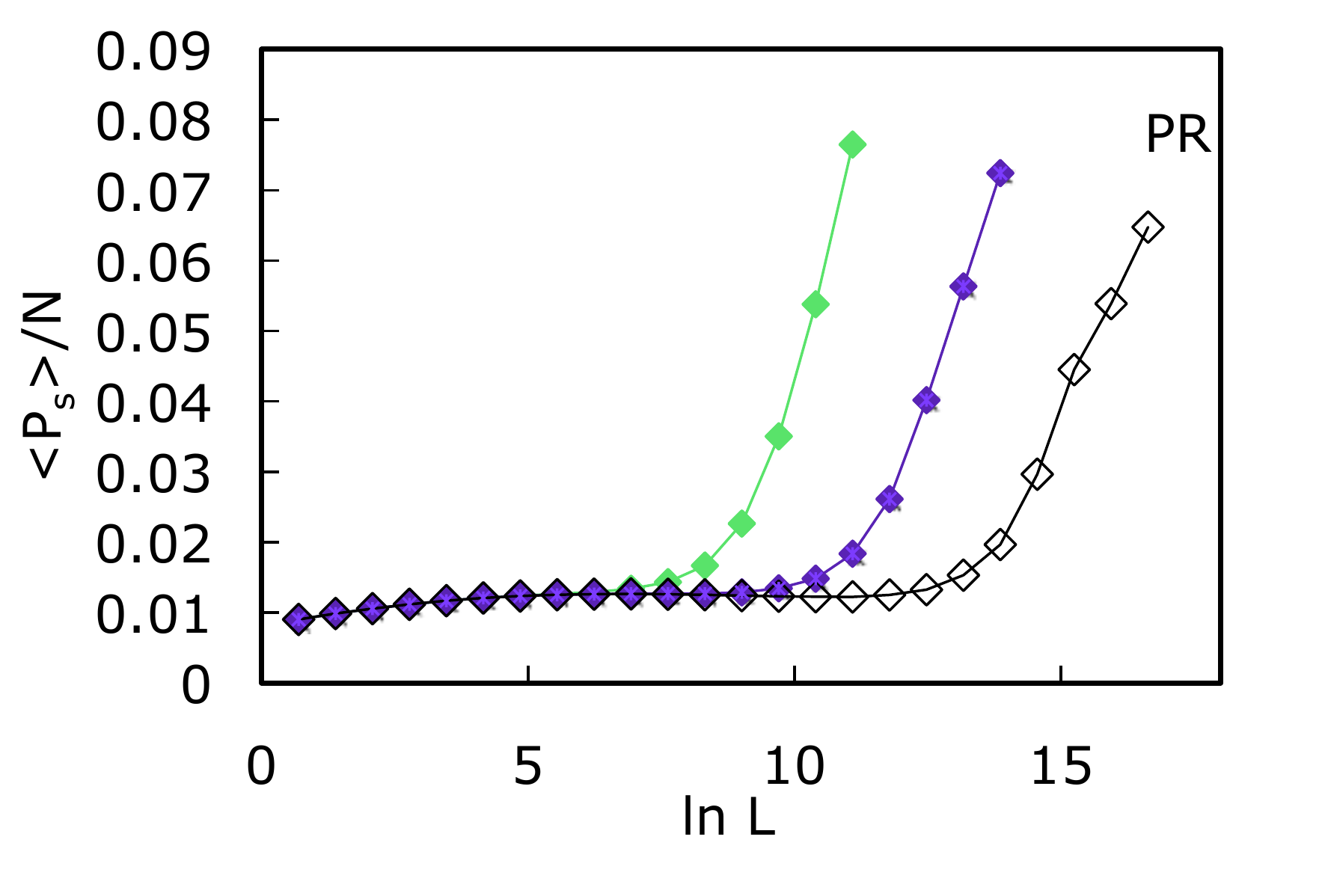} 
   \includegraphics[width=3in]{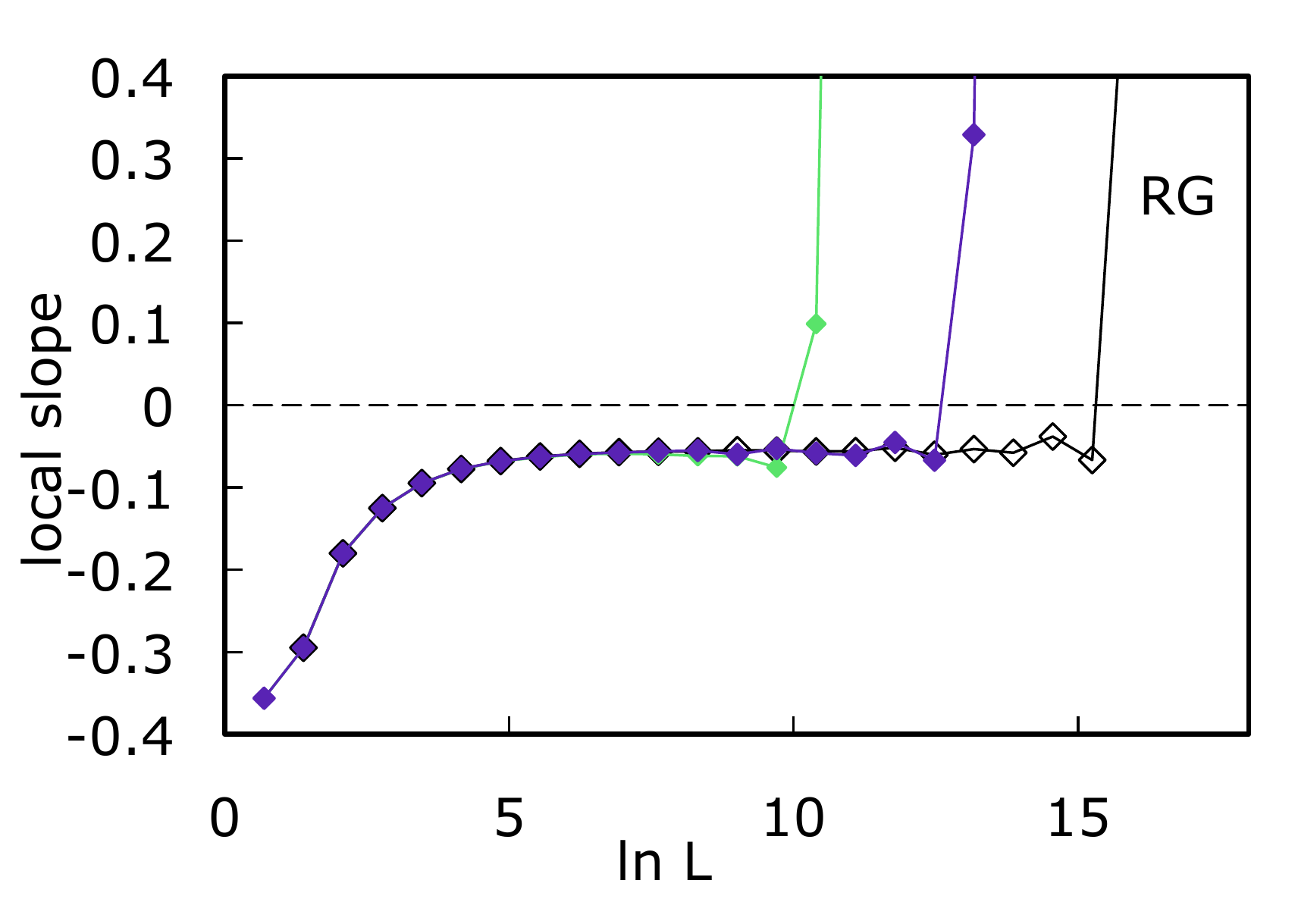} 
   \includegraphics[width=3in]{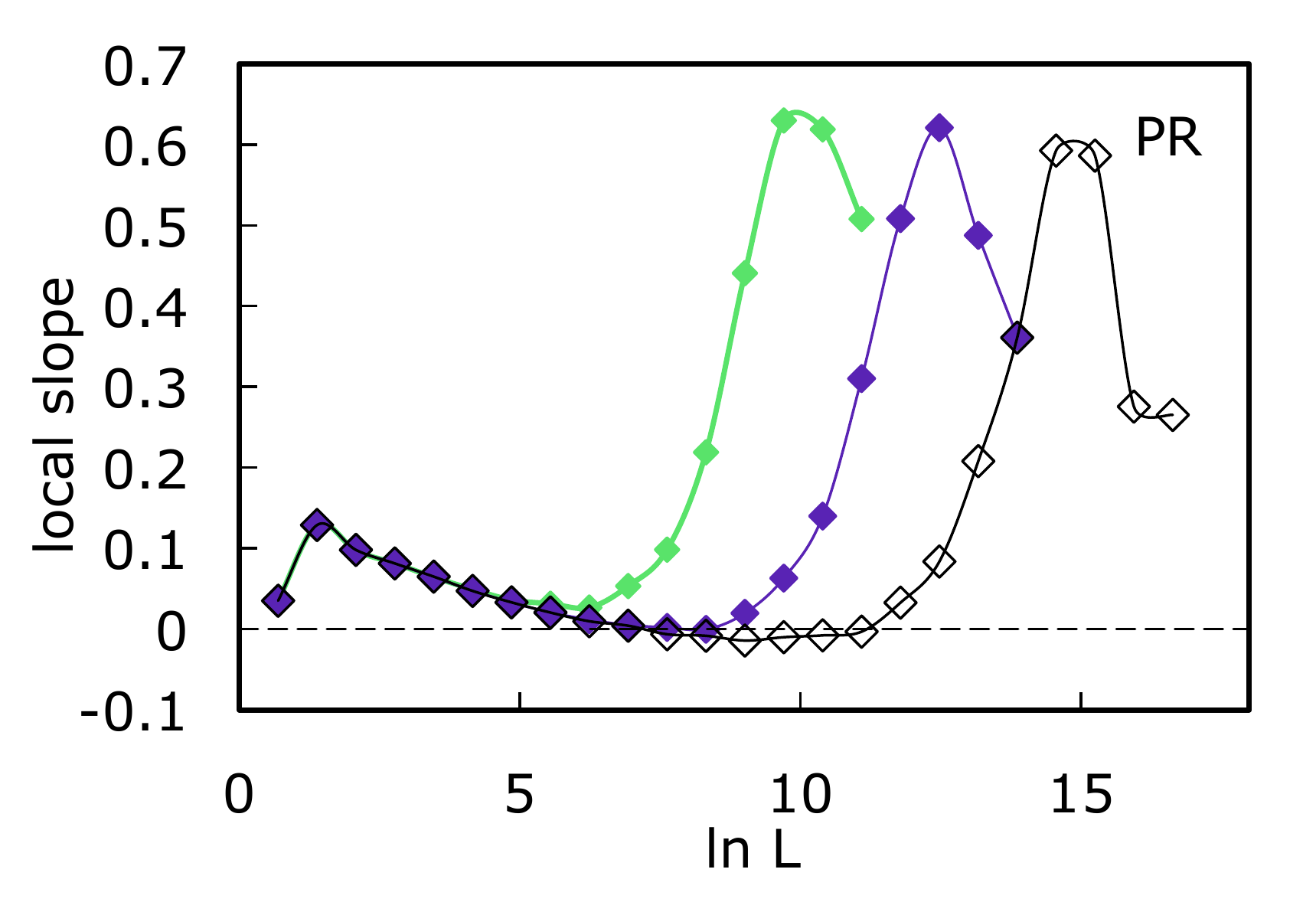} 
   \caption{(Color online) $\langle P_s\rangle / N  $ vs.\ $\ln L$ (upper plots), and logarithmic slopes between pairs of points (lower plots), for $L = 512$, $2048$, and $8192$ (peaks going from left to right).}
   \label{fig:newsizedist}
   \end{figure*}

\subsection{ Size distribution}

Finally, we consider the behavior of the cluster size distribution at criticality.  We ran simulations on systems of size $L = 512$, $2048$, and $8192$ for both the RG and PR models, and measured $n_s$, the number of clusters of size $s$, at the critical points $p_c$.   We binned the weighted data as $P_s = \sum_{s'=s}^{2 s - 1} s' n_{s'}$ for $s = 1, 2, 4, 8, \ldots$, thus accumulating the
number of occupied sites belonging to clusters in each size range.  That is, when growing a cluster of size $s$, we incremented the bin
$n = $ (int)$(\log_2 s)$ by $s$.  
(This gives better statistics than the usual method of incrementing the bin by $1$, which corresponds to just counting the number
of clusters in each size bin.)
For a given run, $\sum_{n\ge 0} P_s = N$ where $s = 2^n$, because, in the end, all $N$ sites are wetted.  
For RG, one expects
 $P_s \sim s^{2 - \tau} \sim s^{-5/91}$, so $P_s$ is a slowly decreasing function of $s$, until $s$ approaches the
 size of the system, at which point the ``infinite" clusters contribute.
 
 In Fig.\ \ref{fig:newsizedist} we plot the average of the normalized distribution, $\langle P_s\rangle/N $, as a function of
 $s$, for different $L$.
For the RG case, the data show expected decrease with $s$, except for a large accumulation in the last two bins because of the finite size effects.  On the
other hand, for the PR case, the $P_s/N$ seem to be increasing, except possibly for a small region in the 
largest system, and the accumulation in the large bins occurs over a much wider range.

In the lower plots of Fig.\ \ref{fig:newsizedist} we show the slopes between pairs of points
(taking the logarithm of $\langle  P_s \rangle /N$ first).
The data of the slopes for the RG model for large $L$ is seen to be consistent with the theoretical prediction $2 - \tau = -5/91$.
For the PR model, for smaller $s$ and $L$, the slopes are positive,
consistent with the observation of \cite{RadicchiFortunato10} who found $\tau = 1.9(1)$.
Of course, for a normalizable size distribution, it is necessary that $\tau > 2$, at least
asymptotically.  We indeed find that the slope (barely) goes below zero (corresponding
to $\tau > 2$) for a range of  $s$ for the largest system; however, it is unclear from these data whether the slope
truly approaches a consistent value or whether it contains for example logarithmic terms. Simulations on larger
systems should help to answer this question.

For the corrections to scaling for the critical size distribution, one expects
\begin{equation}
P_s \sim  s^{2-\tau} (A + B s^{-\Omega}\ldots) \ .
\label{eq:corrections}
\end{equation}
The data for RG are consistent with $\Omega \approx 0.75$ as found previously
\cite{ZiffBabalievski99,AdlerMoshePrivman82}.  If we fit the data
of the PR model to (\ref{eq:corrections}), we find
$B$ is negative, $\Omega \approx 0.3$, and $\tau \approx 2.025$.
The latter value  is consistent with $\beta/\nu = 0.05$ through
the scaling relation $\tau = 2 + \beta/(\nu D)$,
assuming $D = 2$.
The hyper-scaling relation $\beta/\nu = d - D$ 
would imply that $D \approx 1.95$, and the scaling
is also consistent with this value of $D$.
Thus, there is evidence that the size distribution becomes power-law
and that scaling is satisfied for the situations in which $\nu \ne 1$ and $\gamma/\nu \ne 2$.

 \section{Conclusions}

We have found the critical bond fraction $p_c$ for the PR model on the square lattice to high accuracy by a number of methods.
The two best criteria to determine $p_c$ (in terms of convergence with $L$) are the average value of $p$ at which horizontal wrapping first occurs 
(Fig.\ \ref{fig:pcestimates}), and the crossing point of the horizontal wrapping probability (Fig.\ \ref{fig:horiz}).  (One could just as well use vertical wrapping, or the sum of the two \cite{LanglandPouliotSaintAubin94}, as a criterion.)  
Combining our measurements, we conclude
\begin{equation}
 p_c = 0.526565 \pm 0.000005
 \end{equation}
 where the error bars represent a combination of statistical error and also the variation among results based upon
 different criteria.  This
 is consistent with the value 0.5266(2) given by Radicchi and Fortunato \cite{RadicchiFortunato10}.

The striking qualitative different between the explosive and regular percolation 
is highlighted by the non-zero limiting behaviors of $M_2'(p)/N$
(Fig.\ \ref{fig:M2prime}) and $\chi/N$ (Fig.\ \ref{fig:susceptibility}).  These results suggest 
a  discontinuity at the transition point, in contrast to RG 
(regular percolation), where the corresponding quantities are continuous.  (Note that for regular percolation on a hierarchical 
small-world network, however, the transition can also be discontinuous \cite{BoettcherCookZiff09}).

The cluster size distribution of the PR model shows quite different behavior
than the ER model, with possible power-law
behavior for very large systems with strong finite-size effects.

The wrapping probabilities proved  useful for locating the transition point and, perhaps surprisingly, behave  qualitatively 
quite similar to the RG model.   The horizontal wrapping probability $\Pi^{(h)}$ shows a very well-defined crossing point,
just as found for the RG case.  
Its value, $\Pi^{(h)} = 0.5160$, is quite close to (but not identical with) the
value for standard percolation, $\Pi^{(h)} = 0.52105829\ldots $\cite{Pinson94}.   
This result recalls the recent findings of various kinetic systems that evolve to mimic random percolation \cite{BarrosKrapivskyRedner09,SiciliaSarrazinArenzonBrayCugliandolo09}.  

On the other hand, the value of the one-way wrapping probability,
$\Pi^{(1)}$ for the PR model (see Fig.\ \ref{fig:onewaywidth})  is quite a bit below the RG percolation value,
$0.351642855\ldots$, and it is hard to find its asymptotic value precisely.  Evidently, because of the more compact geometry
of the PR giant cluster, wrapping one way is more difficult than in the RG case.

Finally, for the scaling, we have found some contradictory results: $M_2$, $\langle s_\mathrm{max}\rangle$, 
$\Pi^{(1)}$, and the size distribution give $\beta/\nu = 0.06(1)$, $\gamma/\nu = 1.90(1)$,
and $\nu = 1.04(1)$, and $\tau = 2.025(10)$,  implying $D = 1.94(1)$,  where number in parentheses
represents our estimated errors in the last digit(s), while some of the other results (such as the 
behavior of $M_2'$ and $\xi$)
are more consistent with $\nu = 1$ and  $\gamma/\nu = 2.$ \   Perhaps this is indicative that the normal
two-parameter scaling does not hold for this model because of the first-order transition,
or that logarithmic corrections come into play.

{\it Note added:}  While this paper was in revision, a preprint appeared which argues that the explosive percolation transition in the case of the PR rule on the random graph is continuous \cite{CostaDorogovtsevGoltsevMendes10}.  Those arguments however do not apply to the regular square lattice studied here. 

\section*{ACKNOWLEDGMENTS}

The author acknowledges support from the U. S. National Science Foundation grant number DMS-0553487, and 
also acknowledges useful correspondence with R. D'Souza, S. S. Manna, F. Radicchi and S. Fortunato.

\bibliographystyle{apsrev4-1}

\bibliography{PercAchlioptas10PREresubmit.bib}

\end{document}